\documentclass[useAMS,usenatbib]{mn2e}
\usepackage{natbib}
\usepackage{graphicx}
\usepackage{times}
\usepackage{amssymb, amsmath}
\usepackage{epstopdf}
\usepackage{color}
\usepackage[section]{placeins}
\usepackage[english]{babel}
\usepackage{lineno}
\usepackage{url}
\usepackage{ulem}
\usepackage[colorlinks,linkcolor={blue},citecolor={blue},urlcolor={blue}]{hyperref}
\def\mpchi{\,h^{-1}{\rm {Mpc}}}

\def\msun{\,h^{-1}{\rm M_{\sun}}}

\def\aap{A\&A}

\def\apj{ApJ}
\def\apjs{ApJS}
\def\apjl{ApJL}

\def\mnras{MNRAS}
\def\aj{AJ}
\def\nat{Nature}

\def\prd{PhRvD}

\begin{document}

\title[Galaxy Velocity Bias]{Velocity bias from the small-scale clustering of SDSS-III BOSS galaxies}

\author[Guo et al.]{\parbox{\textwidth}{
Hong Guo$^{1}$\thanks{E-mail: hong.guo@utah.edu}, Zheng Zheng$^{1}$, Idit Zehavi$^{2}$, Kyle Dawson$^{1}$, Ramin A. Skibba$^{3}$, Jeremy L. Tinker$^{4}$, David H. Weinberg$^{5,6}$, Martin White$^{7,8,9}$, Donald P. Schneider$^{10,11}$}
\vspace*{6pt} \\
$^{1}$ Department of Physics and Astronomy, University of Utah, UT 84112, USA\\
$^{2}$ Department of Astronomy, Case Western Reserve University, OH 44106, USA\\
$^{3}$ Center for Astrophysics and Space Sciences, University of California,
9500 Gilman Drive, San Diego, CA 92093, USA\\
$^{4}$ Center for Cosmology and Particle Physics, New York University, New York, NY 10003, USA\\
$^{5}$ Department of Astronomy, Ohio State University, Columbus, OH 43210, USA\\
$^{6}$ Center for Cosmology and Astro-Particle Physics, Ohio State University, Columbus, OH 43210, USA\\
$^{7}$ Lawrence Berkeley National Laboratory, 1 Cyclotron Road, Berkeley, CA 94720, USA\\
$^{8}$ Department of Physics, University of California, 366 LeConte Hall, Berkeley, CA 94720, USA\\
$^{9}$ Department of Astronomy, 601 Campbell Hall, University of California at Berkeley, Berkeley, CA 94720, USA\\
$^{10}$ Department of Astronomy and Astrophysics, The Pennsylvania State University, University Park, PA 16802, USA\\
$^{11}$ Institute for Gravitation and the Cosmos, The Pennsylvania State University, University Park, PA 16802, USA}

\maketitle

\begin{abstract}
We present the measurements and modelling of the projected and redshift-space clustering of CMASS galaxies in the Sloan Digital Sky Survey-III Baryon Oscillation Spectroscopic Survey Data Release 11. For a volume-limited luminous red galaxy sample in the redshift range of $0.48<z<0.55$, we perform halo occupation distribution modelling of the small- and intermediate-scale ($0.1$--$60\mpchi$) projected and redshift-space two-point correlation functions, with an accurate model built on high resolution $N$-body simulations. To interpret the measured redshift-space distortions, the distribution of galaxy velocities must differ from that of the dark matter inside haloes of $\sim 10^{13}$--$10^{14}\msun$, i.e. the data require the existence of galaxy velocity bias. Most notably, central galaxies on average are not at rest with respect to the core of their host haloes, but rather move around it with a 1D velocity dispersion of $0.22^{+0.03}_{-0.04}$ times that of the dark matter, implying a spatial offset from the centre at the level of $\lesssim$1 per cent of the halo virial radius. The luminous satellite galaxies move more slowly than the dark matter, with velocities $0.86^{+0.08}_{-0.03}$ times those of the dark matter, which suggests that the velocity and spatial distributions of these satellites cannot both be unbiased. The constraints mainly arise from the Fingers-of-God effect at nonlinear scales and the smoothing to the Kaiser effect in the translinear regime; the robustness of the results is demonstrated by a variety of tests. We discuss the implications of the existence of galaxy velocity bias for investigations of galaxy formation and cosmology.
\end{abstract}

\begin{keywords}
galaxies: distances and redshifts --- galaxies: haloes --- galaxies: statistics --- cosmology: observations --- cosmology: theory --- large-scale structure of Universe
\end{keywords}

\section{Introduction}
Contemporary large galaxy redshift surveys, such as the Sloan Digital Sky Survey-III \citep[SDSS-III;][]{Eisenstein11}, can map the 3D galaxy distribution in great detail. The line-of-sight (LOS) distances of the galaxies from the observer are usually obtained through galaxy redshifts. The observed redshift has two contributions, cosmological redshift from Hubble expansion and Doppler effect from the peculiar velocity of galaxies. The latter is related to the galaxy kinematics in various environments. In this paper, we present constraints on such kinematics from clustering of galaxies in the SDSS-III Baryon Oscillation Spectroscopic Survey (BOSS; \citealt{Dawson13}).

The redshift-space distribution of galaxies is obtained by converting galaxy redshifts to distances under the assumption that the redshift is produced solely from the Hubble expansion. The existence of the galaxy peculiar velocity distorts the galaxy distribution, leading to anisotropy. Such a distortion is reflected in the redshift space as two main effects. On large scales, the infall of galaxies towards overdense regions as well as the streaming of galaxies out of underdense regions compresses their distribution towards overdense regions in the LOS direction; this is known as the Kaiser effect \citep{Kaiser87,Hamilton92} and was recognized in \citet{Sargent77} and \citet{Peebles80}. On small scales, the random motion of galaxies in virialized structures causes the redshift-space distribution of galaxies to appear stretched along the LOS direction, producing the `Fingers-of-God' (FOG) feature \citep{Jackson72,Huchra88}.

The above redshift-space distortion (RSD) effects can be measured in the two-point correlation function (2PCF) of galaxies, $\xi(r_p,r_{\rm\pi})$, where $r_p$ and $r_{\rm\pi}$ are the transverse and LOS separations of galaxy pairs, respectively. In a contour plot of $\xi(r_p,r_{\rm\pi})$, the Kaiser effect causes the contours to be squashed along the LOS separation on large scales, and the FOG effect shows up as elongated contours along the LOS separation at small $r_p$. The FOG effect was clearly detected even in early galaxy surveys \citep[e.g.,][]{Peebles79,Davis83, Bean83,Sadler84,deLapparent88}. As a result of the relatively weak clustering amplitude on large scales, the Kaiser effect was detected later with larger galaxy samples and more accurate redshift measurements \citep[e.g.,][]{Hamilton93b,Cole94a,Fisher94a,Fisher94b}. The two RSD effects have now become common features seen in contemporary galaxy redshift surveys \citep[e.g.,][]{Peacock01,Zehavi02,Zehavi05,Zehavi11,Coil04b,Guzzo08,White11,Guo13}.

The large-scale RSD provides an approach to measure the density of matter that sources the peculiar velocities and the growth rate of structure. To model the RSD, the nonlinear effect has to be taken into account, in addition to the Kaiser effect. A simple streaming model is often used to describe the redshift-space clustering by convolving the real-space clustering with the galaxy pairwise velocity distribution \citep{Peebles80,Peacock94}. The velocity dispersion parameter introduced in such a model intends to capture the velocity dispersion of galaxies in nonlinear, virialized structures. However, such a velocity dispersion at most can be regarded as some sort of average and what it exactly refers to is not clear, since galaxy velocity dispersion depends on environment and scale, which are connected to the mass and size of their host dark matter haloes. One commonly used exponential dispersion ansatz can even lead to a non-physical distribution of pairwise velocities \citep[][]{Scoccimarro04}.

The halo occupation distribution (HOD) framework \citep{Jing98,Peacock00,Seljak00,Scoccimarro01,Berlind02,Berlind03,Yang03,Zheng05} provides a more physical and informative approach to model galaxy clustering, including the RSD effects \citep[e.g.,][]{White01,Yang04,Tinker06,Tinker07,Zu13,Reid14}. The HOD describes the relation between galaxies and dark matter at the level of individual dark matter haloes. It specifies the probability distribution $P(N|M)$ of having $N$ galaxies of a given type in a dark matter halo of virial mass $M$, and the spatial and velocity distributions of these galaxies. The probability distribution $P(N|M)$ and the spatial distribution of galaxies inside haloes, together with the halo population, are sufficient to model the real-space clustering (e.g., the projected 2PCF $w_p(r_p)$). To model the redshift-space clustering, the velocity distribution of galaxies inside haloes is required. We thus can potentially constrain the motion of galaxies inside haloes from measurements of the redshift-space clustering. With the increasing volume of contemporary galaxy redshift surveys, the precision of the data requires more accurate models \citep[e.g.,][]{Reid12,Samushia14} to interpret the redshift-space clustering and extract galaxy kinematics information.

For the velocity distribution of galaxies, theoretical and observational studies have found evidence that central galaxies are not necessarily at rest at the halo centres \citep[e.g.,][]{Berlind03,Yoshikawa03,Bosch05,Skibba11,Li12}. Moreover, subhaloes or satellite galaxies in $N$-body and hydrodynamic simulations of galaxies are found to have velocities differing from those of the dark matter \citep[e.g.,][]{Diemand04,Gao04,Gill04,Lau10,Munari13,Wu13a,Wu13b}. The effect of such velocity bias should appear in redshift-space clustering and be included in detailed modelling of the RSD.

In this paper, building on an accurate simulation-based model, we constrain galaxy kinematics inside dark matter haloes from the redshift-space clustering of the massive galaxies in the CMASS sample of the SDSS-III BOSS. In Section 2, we describe the data and the 2PCF measurements for a volume-limited luminosity threshold galaxy sample, and introduce our modelling method. In Section 3, we present the constraints on the galaxy velocity distribution inside dark matter haloes, which show differences between the galaxies' motion and the dark matter motion, i.e., velocity bias. We perform a variety of tests to establish the robustness of the results, and then discuss the implications of the results for galaxy formation and cosmology in Section 4. Finally, the results are summarized in Section 5. In the appendices, we provide a test of fibre-collision correction with a collision-free sample and provide the constraints on galaxy kinematics for the whole CMASS sample.

Throughout the paper, we assume a spatially flat $\Lambda$ cold dark matter ($\Lambda$CDM) cosmology, with $\Omega_m=0.27$, $h=0.7$, $\Omega_b=0.047$, $n_s=0.95$, and $\sigma_8=0.82$, consistent with the constraints from the seven-year \textit{Wilkinson Microwave Anisotropy Probe} data \citep{Komatsu11}, which are used in the simulation for our model.

\section{Data and Methodology}\label{sec:data}
\subsection{Observations and Measurements}
The SDSS-III BOSS selects galaxies for spectroscopic observations from the five-band SDSS imaging data \citep{Fukugita96,Gunn98,Gunn06,York00}. A detailed overview of the BOSS is given by \cite{Dawson13}, and the spectrographs are described in \cite{Smee13} and \cite{Bolton12}. About $5$ per cent of the fibres are devoted to additional ancillary targets, probing a wide range of objects \citep[see details in][]{Dawson13}.

We focus in this paper on the analysis of a volume-limited BOSS CMASS galaxy sample selected from the SDSS-III BOSS DR11 \citep{Anderson13}. The DR11 CMASS galaxy sample covers an effective area of about $8500\deg^2$ and is designed to select massive galaxies of a typical stellar mass of $10^{11.3}\msun$. We construct a well-defined, volume-limited luminous red galaxy sample, satisfying the following selection criteria \citep{Guo14}
\begin{eqnarray}
(r-i)&>&0.679-0.082(M_i+20),\label{eq:colourcut}\\
M_i&<&-21.6,\\
0.48&<&z\quad<0.55,
\end{eqnarray}
where the Galactic extinction-corrected \citep{Schlegel98} absolute magnitude $M_i$ and $r-i$ colour are both $k+e$ corrected to $z=0.55$ \citep{Tojeiro12}. CMASS galaxies have a complex sample selection \citep{Eisenstein11}. The colour cut of Equation (\ref{eq:colourcut}) is intended to select red galaxies \citep[see, e.g., fig. 1 in][]{Guo13}. This luminosity-threshold sample covers a volume of $V_{\rm obs}=0.78\,h^{-3}{\rm Gpc}^3$. With the complex sample selection, the clustering of the full CMASS sample, which is not volume-limited, is not straightforward to model, but given its key role in constraining cosmological parameters \citep[e.g.,][]{Anderson12,Anderson13}, we also present in Appendix~\ref{app:full} the modelling results of the full CMASS sample in the redshift range of $0.43<z<0.7$.

When performing the 2PCF measurements, we follow \cite{Guo14} and employ the fibre-collision correction method of \cite{Guo12}. We also make use of a collision-free galaxy sample from one of the ancillary programme of the BOSS to further test the correction method and confirm that it is working well (see Appendix~\ref{app:fiber}).

We measure the redshift-space 3D 2PCF $\xi(r_p,r_{\rm\pi})$ using the Landy--Szalay estimator \citep{Landy93}. To produce a quantity less affected by RSD effects, we project the 2PCF along the LOS direction to obtain the projected 2PCF $w_p(r_p)$ \citep{Davis83}, which is defined as
%%%%%%%%%%%%%%%%%%%%%%%%%%%%%%%%%%%%%%%%%%%%%%%%%%
\begin{eqnarray}
w_p(r_p)=2\int_0^\infty \xi(r_p,r_{\rm\pi})dr_{\rm\pi} = 2\sum_i\xi(r_{p},r_{\pi,i})\Delta r_{\pi},\label{eq:wp}
\end{eqnarray}
%%%%%%%%%%%%%%%%%%%%%%%%%%%%%%%%%%%%%%%%%%%%%%%%%%
where $r_{\pi,i}$ and $\Delta r_{\pi}$ are the $i$--th bin of the LOS separation and its corresponding bin size. We set $r_p$ in logarithmic bins centred at $0.13$ to $51.5\mpchi$ with $\Delta\log r_p=0.2$, and $r_{\rm\pi}$ in linear bins from 0 to 100$\mpchi$ with $\Delta r_{\rm\pi}=2\mpchi$. The sum of $\xi(r_p,r_{\rm\pi})$ along the LOS direction is formed up to $r_{\pi, {\rm max}}=100\mpchi$, which is large enough to include most of the correlated pairs.

We further measure the redshift-space 2PCF in bins of $s$ and $\mu$, where $s^2=r_p^2+r_{\rm\pi}^2$ and $\mu$ is the cosine of the angle between $s$ and the LOS direction. Following \cite{Hamilton92}, the redshift-space 2PCF $\xi(s,\mu)$ can be written in the form of a multipole expansion,
%%%%%%%%%%%%%%%%%%%%%%%%%%%%%%%%%%%%%%%%%%%%%%%%%%
\begin{equation}
\xi(s,\mu)=\sum_l\xi_l(s)P_l(\mu),
\end{equation}
%%%%%%%%%%%%%%%%%%%%%%%%%%%%%%%%%%%%%%%%%%%%%%%%%%
where $P_l$ is the $l$-th order Legendre polynomial and the multipole moment $\xi_l$ is calculated through
%%%%%%%%%%%%%%%%%%%%%%%%%%%%%%%%%%%%%%%%%%%%%%%%%%
\begin{eqnarray}
\xi_l(s) & = & \frac{2l+1}{2}\int_{-1}^{1}\xi(s,\mu)P_l(\mu)d\mu\cr
         & = & \frac{2l+1}{2}\sum_i\xi(s,\mu_i)P_l(\mu_i)\Delta\mu.
\label{eq:pole}
\end{eqnarray}
%%%%%%%%%%%%%%%%%%%%%%%%%%%%%%%%%%%%%%%%%%%%%%%%%%
The last equality is how we compute the multipole moments, with $\mu_i$ and $\Delta\mu$ the $i$-th $\mu$ bin and bin width. Similar to $r_p$, we set $s$ in logarithmic bins centred at $0.13$ to $51.5\mpchi$ with $\Delta\log s=0.2$. For $\mu$, we use linear bins from -1 to 1 with $\Delta \mu=0.05$.

In linear theory, only the monopole ($\xi_0$), quadrupole ($\xi_2$), and hexadecapole ($\xi_4$) are non-zero. In this paper, we perform joint modelling of $w_p$, $\xi_0$, $\xi_2$, and $\xi_4$ to constrain the HOD, including the mean occupation function and velocity distribution of galaxies inside haloes.

\subsection{Simulation and Modelling Method}

\subsubsection{The Simulation-Based Model}

To ensure an accurate HOD modelling of the clustering measurements, we adopt a simulation-based model, presented in detail in Zheng \& Guo (in preparation). In brief, to compute galaxy 2PCFs, we tabulate all necessary elements of dark matter haloes in a simulation in fine mass bins ($\Delta\log M =0.01$ in this paper) and adopt the same binning scheme as in the measurements. Specifically, we tabulate different 2PCF components from particles representing the one-halo central-central, central-satellite pairs and the two-halo central-central, central-satellite, and satellite-satellite pairs. The model 2PCF for the galaxies is then obtained as the sum of these 2PCF components weighted by the corresponding mean occupation function from the HOD prescription. This method is accurate and fast in computing the galaxy 2PCF, because it automatically and accurately accounts for the effects halo exclusion, nonlinear growth, and scale-dependent halo bias. This simulation-based model is equivalent to creating mock galaxy samples by populating the haloes in the simulation with galaxies for a given HOD prescription, while it is way faster in computing the 2PCF. Detailed explanation and tests of this method can be found in Zheng \& Guo (in preparation).

In this paper, we build our model based on the MultiDark run simulation (MDR1;\citealt{Prada12,Riebe13}). We use its output at $z=0.53$ to match the redshift of CMASS galaxies. This $N$-body $\Lambda$CDM cosmological simulation adopts the following cosmological parameters: $\Omega_m=0.27$, $\Omega_b=0.047$, $h=0.7$, $n_s=0.95$, and $\sigma_8=0.82$. The simulation has 2048$^{3}$ dark matter particles in a box of 1\,$h^{-1}$\,Gpc on a side. The mass of the dark matter particle is $8.72\times10^9$ $h^{-1}$ M$_{\sun}$. In our fiducial model, we use dark matter haloes identified with the spherical overdensity (SO) algorithm (Bound Density Maximum technique as described in \citealt{Klypin97}) in MDR1. The SO haloes are defined to have a mean density $\Delta_{\rm vir}$ times that of the background universe, where $\Delta_{\rm vir}\simeq237$ at $z=0.53$ for MDR1 \citep{Bryan98}.

For each halo, we define its centre (the position to place a central galaxy; see below) as the position of the dark matter particle with the minimal potential. We also need to define a characteristic halo velocity. The velocity corresponding to the minimal potential position is not well determined, since the mean velocity of dark matter particles around it depends on radius \citep[e.g.][]{Behroozi13,Reid14}. We therefore adopt a velocity for the
core of the halo, determined from the bulk velocity of the inner $25$ per cent of halo
particles around the above centre. Using the inner $25$ per cent of the halo particles is representative of the halo centre, while relatively insensitive to resolution effects of the simulation used. The velocity of the central galaxy will be formulated with respect to this core velocity, and can also be translated to the one relative to the centre-of-mass halo velocity (corresponding to the bulk velocity of $100$ per cent halo particles). Although the centre-of-mass velocity is commonly used to characterize the halo velocity, the core velocity is more informative as it relates to the distribution of particles in the inner part of a halo, which acts as the immediate environment of the central galaxy. We note that our definition of the halo velocity differs from that of \cite{Reid14}, but when converted to the same definition the results on velocity bias are consistent (see details in Appendix \ref{app:full}).  To investigate the effect of halo definition in our analysis, we also present modelling results in a few cases for haloes identified by a friends-of-friends (FOF) algorithm with a linking length of $0.17$ in MDR1. For further details on the MultiDark simulation, we refer the readers to \citet{Prada12}.

Since the measurements are all produced from redshift-space 2PCFs, we build our simulation tables in redshift space as well. Given the narrow redshift range considered in this paper, we adopt the plane-parallel approximation and use the $\hat{z}$ direction in the simulation as the LOS. In redshift space, the shift (in comoving units) in the position of a galaxy particle in the $\hat{z}$ direction is calculated as $\Delta Z=v_{\hat{z}}(1+z)/H$, where $v_{\hat{z}}$ is the LOS peculiar velocity of the particle, and $H$ is the Hubble constant at redshift $z=0.53$. The peculiar velocity in our table includes the effect of velocity bias (see below). The simulation tables have all the one-halo and two-halo components of $w_p$, $\xi_0$, $\xi_2$, and $\xi_4$.

\subsubsection{The HOD parametrization}

For the HOD prescription, we separate the contribution to the occupation number into those from central galaxies and satellites \citep{Kravtsov04,Zheng05}. We follow the HOD parametrization of \cite*{Zheng07} for the central and satellite mean occupation functions, $\langle N_{\rm cen}(M)\rangle$ and $\langle N_{\rm sat}(M)\rangle$,
%%%%%%%%%%%%%%%%%%%%%%%%%%%%%%%%%%%%%%%%%%%%%%%%%%
\begin{eqnarray}
\langle N_{\rm cen}(M)\rangle&=&\frac{1}{2}\left[1+\rm{erf}\left(\frac{\log M-\log M_{\rm min}}{\sigma_{\log
M}}\right)\right], \label{eqn:Ncen}\\
\langle N_{\rm sat}(M)\rangle&=&\langle N_{\rm cen}(M)\rangle\left(\frac{M-M_0}{M_1^\prime}\right)^\alpha, \label{eqn:Nsat}
\end{eqnarray}
%%%%%%%%%%%%%%%%%%%%%%%%%%%%%%%%%%%%%%%%%%%%%%%%%%
where ${\rm erf}$ is the error function, $M_{\rm min}$ describes the cutoff halo mass of the central galaxies and $\sigma_{\log M}$ is the width of the cutoff profile. The three parameters for the satellite galaxies are the cutoff mass scale $M_0$, the normalization mass scale $M_1^\prime$ and the high-mass slope $\alpha$ of $\langle N_{\rm sat}(M)\rangle$.

In the simulation-based model, in each halo hosting a galaxy in our sample we assign the central galaxy to reside at the centre of the halo (defined as the position of potential minimum). For satellite galaxies, in our fiducial model, we assign the positions of randomly selected dark matter particles from the halo. Our model preserves the non-spherical shapes of haloes. The assumption that the spatial distribution of satellite galaxies inside haloes follows that of the dark matter is tested and discussed in Section \ref{sec:sys}.

To assign velocities to central and satellite galaxies, we introduce two additional HOD parameters to account for the possible differences in the velocity distributions of galaxies and dark matter inside haloes (a.k.a. velocity bias). The two velocity bias parameters, $\alpha_c$ and $\alpha_s$, apply to central and satellite galaxies, respectively.

For each halo in the simulation, we measure the 1D velocity dispersion $\sigma_v$ of dark matter particles. The central galaxy particle (at the potential minimum) is then assigned a LOS velocity (with respect to the core velocity of the halo) following a Gaussian distribution with zero mean and standard deviation of
\begin{equation}
\label{eqn:alpha_c} \sigma_c=\alpha_c\sigma_v.
\end{equation}
This accounts for the possibility that the central galaxy may not be at rest with respect to the core.

To incorporate the velocity bias for satellite galaxies, the LOS velocity $v_s$ of each satellite is assigned through
\begin{equation}
\label{eqn:alpha_s} v_s-v_h=\alpha_s(v_p-v_h),
\end{equation}
where $v_h$ is the LOS velocity of the halo core and $v_p$ is the original LOS velocity of the dark matter particle that represents the galaxy, i.e. in the rest frame of the halo core, the velocity of the satellite galaxy differs from that of the dark matter particle by a factor of $\alpha_s$. And the consequence is that the 1D velocity dispersion $\sigma_s$ of satellite galaxies is a factor of $\alpha_s$ times that of the dark matter particles, $\sigma_s=\alpha_s\sigma_v$ \citep[see, e.g.,][]{Tinker07}.

In addition to the velocity related to the velocity bias, we also incorporate into the model the dispersion caused by the measurement error in the redshifts of CMASS galaxies. From galaxies with repeated observations, we find that the root-mean-square (rms) redshift error corresponds to a LOS velocity dispersion of $30 {\rm km\, s^{-1}}$ (see also \citealt{Bolton12}). Therefore, in building our simulation tables, we add to each (central or satellite) galaxy particle a LOS velocity, drawing from a Gaussian distribution with a zero mean and a standard deviation of $30\,{\rm km\, s^{-1}}$.

We emphasize that even though we phrase the velocity bias as a modification to the LOS velocities (Equations~\ref{eqn:alpha_c} and \ref{eqn:alpha_s}), it applies to all the three velocity components, i.e. the velocity bias is for the 3D velocity of galaxies. For the purpose of modelling redshift-space clustering, however, only the LOS component matters.

The velocity bias effects have been incorporated in the simulation tables in our model. We construct tables for $\alpha_c$ in the range of $[0,1]$ and $\alpha_s$ in $[0,2]$ in bins of $\Delta\alpha_{c,s}=0.1$. The model interpolates among these tables for any given values of $\alpha_c$ and $\alpha_s$. We have verified that using smaller bin sizes for the interpolation does not produce any significant difference in the results. In total, our fiducial model has seven free HOD parameters, five for the mean occupation function and two for the velocity bias. The two velocity bias parameters are the key parameters we aim to constrain in this paper. A model with no velocity bias corresponds to $\alpha_c = 0$ and $\alpha_s = 1$. As shown in the next section, the measurements in CMASS galaxies in fact require substantial velocity bias.

\subsubsection{Modelling the 2PCF Measurements}
\begin{figure}
\includegraphics[width=0.49\textwidth]{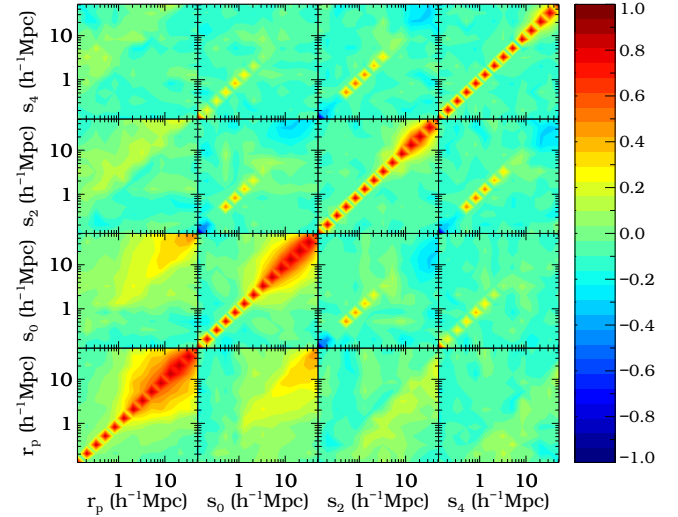}\caption{Normalized jackknife error covariance matrix for the
volume-limited, luminosity-threshold ($M_i<-21.6$) sample. Different panels show the covariance for the measurements of $w_p(r_p)$, $\xi_0(s_0)$, $\xi_2(s_2)$, and $\xi_4(s_4)$ (from left to right and bottom to up).} \label{fig:cov}
\end{figure}
In our modelling of the clustering measurements, we apply a Markov chain Monte Carlo method to explore the HOD parameter space. We jointly fit the projected 2PCF $w_p(r_p)$, the redshift-space multipoles $\xi_0(s)$, $\xi_2(s)$ and $\xi_4(s)$, and the observed number density $n_g$ of galaxies in the sample. The $\chi^2$ of the fitting is defined as
%%%%%%%%%%%%%%%%%%%%%%%%%%%%%%%%%%%%%%%%%%%%%%%%%%
\begin{equation}
\chi^2= \bmath{(\xi-\xi^*)^T C^{-1} (\xi-\xi^*)}
       +\frac{(n_g-n_g^*)^2}{\sigma_{n_g}^2}, \label{eq:chi2}
\end{equation}
%%%%%%%%%%%%%%%%%%%%%%%%%%%%%%%%%%%%%%%%%%%%%%%%%%
where $\bmath{C}$ is the full error covariance matrix and the data vector $\bmath{\xi} = [\bmath{w_p},\bmath{\xi_0},\bmath{\xi_2},\bmath{\xi_4}]$, i.e. the combination of the measurements of the projected 2PCF and redshift-space 2PCF multipoles. The quantity with (without) a superscript `$*$' is the one from the measurement (model).
\begin{table}
\caption{HOD parameters and satellite fraction derived for the volume-limited $M_i<-21.6$ sample} \label{tab:hod}
\begin{tabular}{lrr}
\hline
Parameters     & SO halo model             & FOF halo model\\
\hline
$\chi^2/\rm{dof}$  & $55.91/50$     & $61.16/50$ \\
$\log M_{\rm min}$ & $13.36^{+0.03}_{-0.05}$ & $13.36^{+0.04}_{-0.03}$ \\
$\sigma_{\log M}$  & $0.65^{+0.03}_{-0.04}$  & $0.67^{+0.03}_{-0.03}$ \\
$\log M_0$         & $13.28^{+0.20}_{-0.45}$ & $13.60^{+0.15}_{-0.39}$\\
$\log M_1^\prime$  & $14.21^{+0.07}_{-0.04}$ & $13.98^{+0.19}_{-0.17}$\\
$\alpha$           & $1.02^{+0.30}_{-0.35}$  & $1.02^{+0.40}_{-0.30}$\\
$\alpha_c$         & $0.22^{+0.03}_{-0.04}$  & $0.23^{+0.03}_{-0.02}$\\
$\alpha_s$         & $0.86^{+0.08}_{-0.03}$  & $0.69^{+0.03}_{-0.04}$\\
$f_{\rm{sat}}(\rm{per\,cent})$ & $6.96^{+0.46}_{-0.32}$  & $7.13^{+0.28}_{-0.46}$\\
\hline
\end{tabular}

\medskip
The mean number density of the sample $\bar{n}(z)$ is $2.19\times10^{-4}h^{3}{\rm {Mpc}}^{-3}$. The halo mass is in units of $\msun$. The best-fitting $\chi^2$ and the dof with the HOD modelling are also given. The degrees of freedom are calculated as ${\rm dof}=N_{\rm 2PCF}+1-N_{\rm par}$, where the total number of data points ($N_{\rm 2PCF}+1$) is that of the 2PCF data points ($w_p$, $\xi_0$, $\xi_2$, and $\xi_4$) plus one number density constraint, and $N_{\rm par}$ is the number of HOD parameters.
\end{table}
\begin{figure*}
\includegraphics[width=0.7\textwidth]{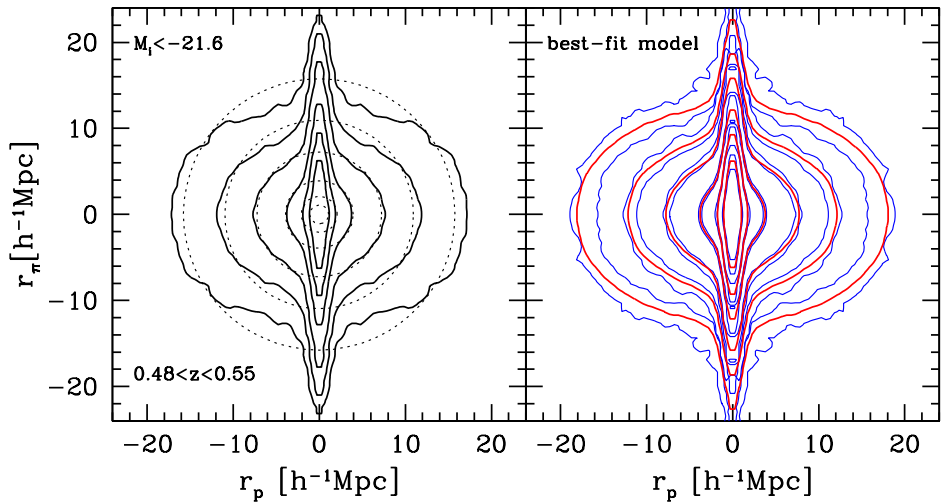}
\caption{ Left: measurements of the 3D 2PCF $\xi(r_p,r_{\rm\pi})$ for the volume-limited $M_i<-21.6$ CMASS sample. Contour levels shown are $\xi(r_p,r_{\rm\pi})=[0.5, 1, 2, 5, 10, 20]$. The dotted curves are the angle-averaged redshift-space correlation function, $\xi_0(s)$, for the same sample and with the same contour levels. Right: the 2PCF from the best-fitting model. The thick red contours are the $\xi(r_p,r_{\rm\pi})$ predicted by our best-fitting HOD model. The pair of thin blue contours around each thick red contour are from the $\pm 1\sigma$ range of the measured $\xi(r_p,r_{\rm\pi})$ to indicate the size of the uncertainties in the measured 2PCF. We do not model $\xi(r_p,r_{\rm\pi})$ directly; this panel serves as a consistency check (see the text). } \label{fig:xip}
\end{figure*}

The covariance matrix is first determined from 403 jackknife subsamples \citep{Zehavi05,Guo13}. We then apply the mean correction for the bias effect described in \cite{Hartlap07} \citep[see also][]{Percival14}. Since the MDR1 simulation has a finite volume $V_{\rm sim}=1\,h^{-3}{\rm Gpc}^3$, only slightly larger than the volume $V_{\rm obs}=0.78\,h^{-3}{\rm Gpc}^3$ of the observational sample, the model uncertainty is of significance. We incorporate the model uncertainty by rescaling the above covariance matrix by a factor of $1+V_{\rm obs}/V_{\rm sim}=1.78$ (for details, see Zheng \& Guo, in preparation) to obtain the covariance matrix  $\bmath{C}$ used in the modelling. The error $\sigma_{n_g}$ on the number density is determined from the variation of $n_g(z)$ in the different jackknife subsamples.

Fig.~\ref{fig:cov} shows the normalized covariance matrix for the volume-limited $M_i < -21.6$ sample. The data points for each set of measurements ($w_p$, $\xi_0$, $\xi_2$, or $\xi_4$) are positively correlated on large scales, reflecting the effect of sample variance. There are positive correlations between $w_p$ and any one of the multipoles for $s>r_p$. This result is easy to understand by noticing that galaxy pairs at a given separation $r_p$ contribute to $\xi_{0,2,4}(s)$ for $s>r_p$ (as $s^2=r_p^2+r_{\rm\pi}^2$). The correlation becomes weaker between $w_p$ and higher multipoles. Among the multipoles, there are clear positive correlations on scales of a few $h^{-1}{\rm Mpc}$, which should be mainly caused by the FOG effect.

\section{The Modelling Results}\label{sec:results}
\subsection{The Main Results}\label{sec:main_results}
\begin{figure*}
\includegraphics[width=0.7\textwidth]{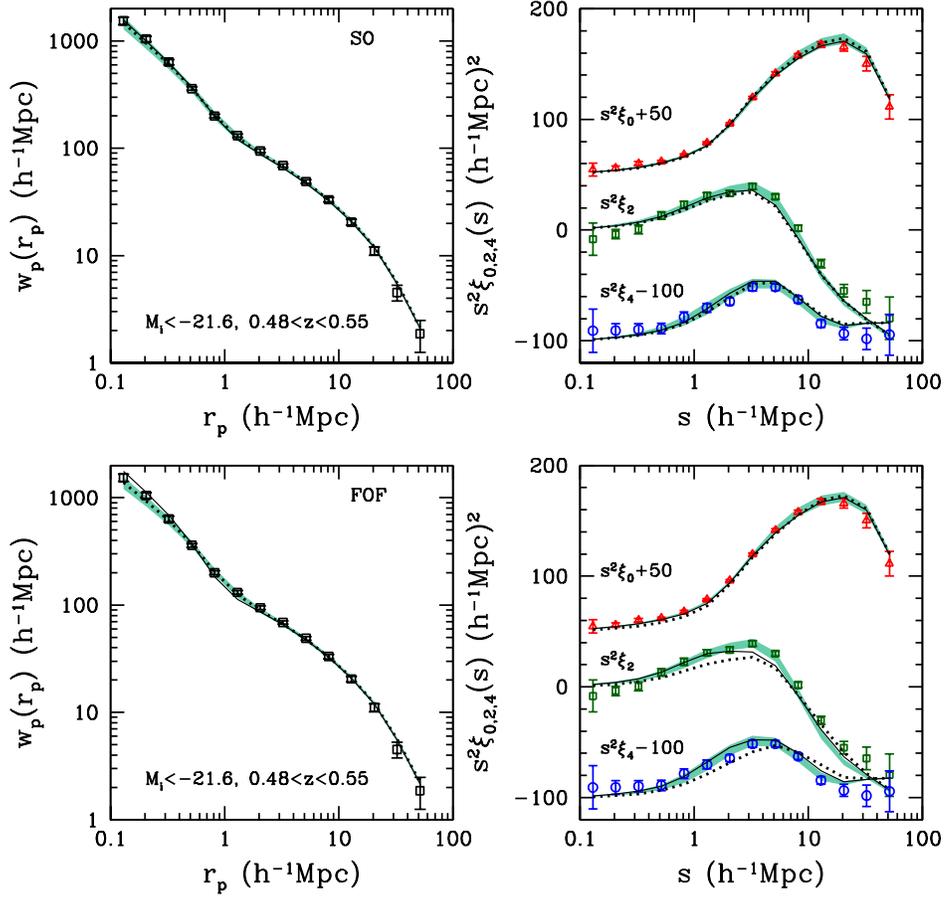}\caption{Left-hand panels:
measurements and best-fitting models of $w_p(r_p)$, with the shaded area showing the $1\sigma$ deviation of the HOD model predictions. The squares are the measurements from the volume-limited $M_i<-21.6$ sample. Right-hand panels: measurements and best-fitting models of the redshift-space 2PCF multipoles, $\xi_0(s)$ (triangles), $\xi_2(s)$ (squares), and $\xi_4(s)$ (circles), respectively. The measurements for $\xi_0(s)$ and $\xi_4(s)$ are vertically shifted for clarity. The shaded areas represent the $1\sigma$ uncertainty in the HOD model predictions. The top and bottom panels are for models with SO haloes and FOF haloes, respectively. The dotted curves in each panel denote the predictions from the best-fitting HOD parameters in Table~\ref{tab:hod}, but with $\alpha_{\rm c}=0$ and $\alpha_s=1$. The solid curves represent the results by fitting the data without any velocity bias parameters included. } \label{fig:wpxi}
\end{figure*}
We first present, in the left-hand panel of Fig.~\ref{fig:xip}, the 3D 2PCF $\xi(r_p,r_{\rm\pi})$ measured for the volume-limited $M_i<-21.6$ sample. The RSD features of the small-scale FOG and large-scale Kaiser squashing effects are clearly seen, indicating the significant peculiar velocities of the CMASS galaxies. We do not model $\xi(r_p,r_{\rm\pi})$ directly, but show in the right-hand panel the {\it predicted} $\xi(r_p,r_{\rm\pi})$ in red from our best-fitting HOD model. The pair of blue contours around each red one are the $\pm 1\sigma$ range of the {\it measured} $\xi(r_p,r_{\rm\pi})$, based on the rescaled jackknife error bars in each $(r_p,r_{\rm\pi})$ cell. The blue contours provide a sense of the uncertainties in the measured $\xi(r_p,r_{\rm\pi})$. This consistency check clearly shows that the best-fitting model well reproduces the measured $\xi(r_p,r_{\rm\pi})$.

We now present our results from simultaneously fitting $w_p$ and the multipoles $\xi_0$/$\xi_2$/$\xi_4$. Fig.~\ref{fig:wpxi} shows our measurements for the projected 2PCF $w_p(r_p)$ (left-hand panels) and the three multipole moments (right-hand panels) for the volume-limited $M_i<-21.6$ sample. The predictions of these quantities from the $1\sigma$ distribution around the best-fitting model to these four sets of measurements are indicated as the shaded areas, for the SO haloes (top panels) and FOF haloes (bottom panels). The best-fitting model parameters are listed in Table~\ref{tab:hod}, including the corresponding $\chi^2$ values. Both models based on the SO and FOF haloes provide reasonably good fits to the data, with a slightly better fit using the SO haloes.  Additional fits (shown as solid and dotted lines) will be discussed later in this section.

\begin{figure*}
\includegraphics[width=0.7\textwidth]{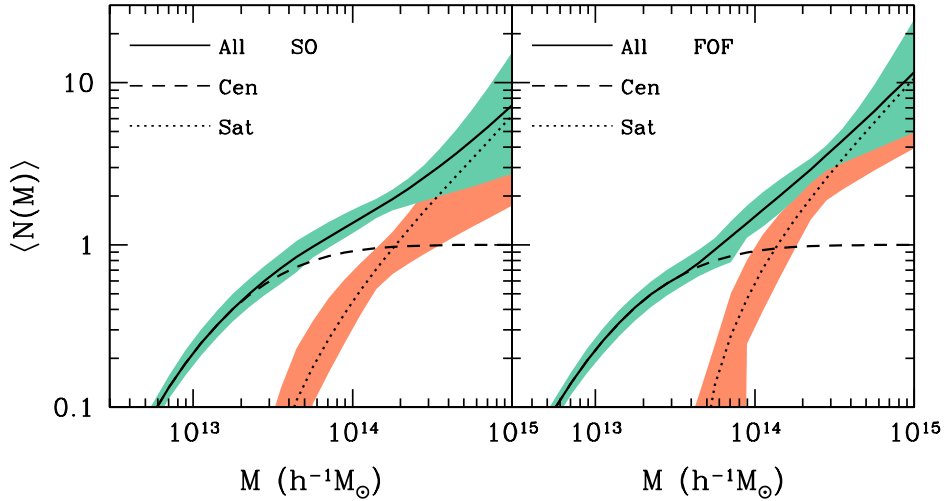}\caption{Mean occupation
functions from the best-fitting models with SO (left-hand panel) and FOF haloes (right-hand panel). In each case, the total mean occupation function (solid curve) is decomposed into contributions from central galaxies (dashed curve) and satellite galaxies (dotted curve). The shaded areas show the $1\sigma$ range around the best-fitting model.} \label{fig:hod}
\end{figure*}
Fig.~\ref{fig:hod} displays the mean occupation functions from the best-fitting models based on SO haloes (left) and FOF haloes (right). In each case, the mean occupation function is decomposed into the contributions from central and satellite galaxies, represented by dashed and dotted curves, respectively. The shaded areas show the $1\sigma$ distribution around the best-fitting model. Compared to the SO model, the FOF model predicts a steeper high-mass end slope $\alpha$ and a higher halo mass cutoff $\log M_0$ for satellite galaxies. The notorious bridging effect in FOF haloes (e.g., \citealt{Tinker08b}) could contribute to such differences, and we discuss this issue further below. The satellite fractions from the two models are quite similar, around $7$ per cent (see Table~\ref{tab:hod}). Both best-fitting models have non-negligible velocity bias parameters. The best-fitting SO model has $\alpha_c=0.22^{+0.03}_{-0.04}$ and $\alpha_s=0.86^{+0.08}_{-0.03}$, while the best-fitting FOF model has $\alpha_c=0.23^{+0.03}_{-0.02}$ and $\alpha_s=0.69^{+0.03}_{-0.04}$. The existence of velocity bias is the main result of this paper.

\begin{figure}
\includegraphics[width=0.47\textwidth]{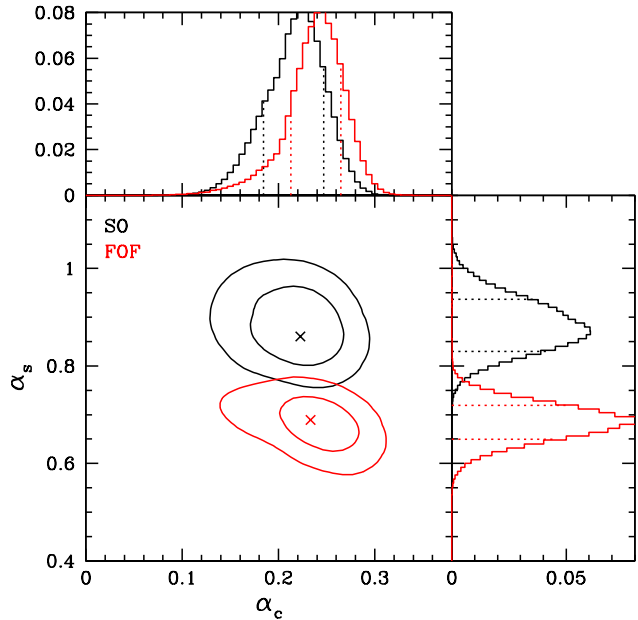}
\caption{ Marginalized probability distributions of the central and satellite velocity bias parameters $\alpha_c$ and $\alpha_s$. The results from the SO and FOF models are in black and red, respectively. The contours show the $68$ and $95$ per cent confidence levels for the two parameters. The crosses correspond to those from the best-fitting models. The marginalized 1D distribution of $\alpha_c$ and $\alpha_s$ are shown in the top-left and bottom-right panels, with the central $68$ per cent distribution indicated by the dotted lines. } \label{fig:alphacs}
\end{figure}

The marginalized joint probability distribution of the velocity bias parameters $\alpha_c$ and $\alpha_s$ is presented in Fig.~\ref{fig:alphacs}. The marginalized 1D probability distributions of $\alpha_c$ and $\alpha_s$ are also shown in the top-left and bottom-right panels, with the central $68.3$ per cent confidence levels indicated by the dotted lines. The FOF model has slightly larger $\alpha_c$ and significantly smaller $\alpha_s$ than the SO model, which may point to the bridging effect for FOF haloes. Without the bridging effect, the distinct haloes in the two halo definitions should have almost the same centres (defined to be the position of potential minimum). When two haloes are bridged together to form one FOF halo, the satellite distribution in the halo would be significantly different from that of the SO model. This effect is more severe considering the fact that the CMASS galaxies mostly occupy massive haloes of $10^{13}$--$10^{14}\msun$ \citep{Guo14}, while many of such massive FOF haloes are actually two haloes bridged together.

From the marginalized 1D probability distributions, we conclude that in the SO model, the existence of satellite velocity bias is at the 1.9$\sigma$ level, while that of the central galaxy velocity bias is at the 5.9$\sigma$ level. In the FOF model, the above numbers become 10.4$\sigma$ and 11.1$\sigma$, respectively. The result implies that on average the central galaxies are not at rest with respect to the core of the halo. Their 1D rms velocity is about 20 to 25 per cent of the velocity dispersion of halo particles. In other words, the mean specific kinetic (``thermal'') energy of central galaxies is about $\sim 4.5$ per cent that of halo particles. The satellite galaxies move on average more slowly than dark matter particles (i.e., they are cooler than dark matter particles). We test the robustness of the results in Section~\ref{sec:sys} and discuss the implications in Section~\ref{sec:implications}. Here we continue with an investigation on the source of the velocity bias constraints.

To demonstrate the effect of velocity bias on the 2PCFs, we compare the best-fitting model with two additional cases in Fig.~\ref{fig:wpxi}. We first display the predictions of the best-fitting HOD parameters in Table~\ref{tab:hod}, but set $\alpha_c=0$ and $\alpha_s=1$. The results are shown as dotted curves in each panel. The corresponding value of $\chi^2$ increases significantly, to $90.5$ for the SO model and $229.1$ for the FOF model. With respect to the best-fitting model, the largest deviation arises in $\xi_2(s)$, around a few Mpc. The small error bars in this range cause a large change in the $\chi^2$ when setting $\alpha_c=0$ and $\alpha_s=1$.

\begin{figure}
\includegraphics[width=0.45\textwidth]{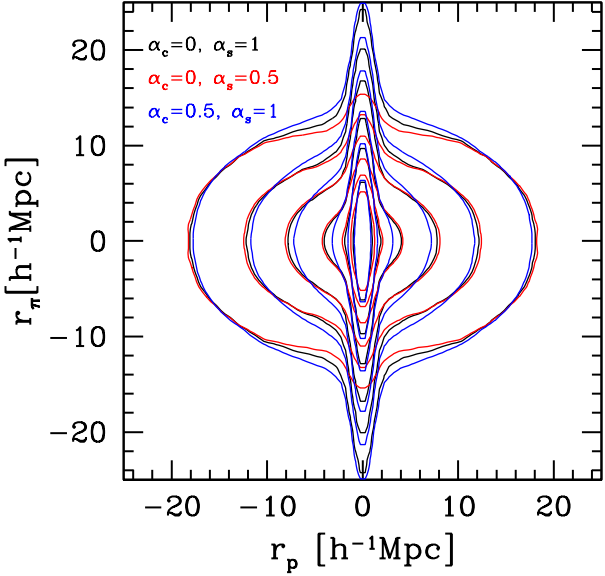}
\caption{An illustration of the effects of velocity bias on the redshift-space 3D 2PCF $\xi(r_p,r_{\rm\pi})$. The black contours are the HOD predictions for the case of no velocity bias. The red contours correspond to a case with only having satellite velocity bias, and the blue contours correspond to having only central velocity bias. A less-than-unity satellite velocity bias makes the FOG feature less elongated, while a central velocity bias makes the Kaiser feature less squashed. Both biases have the effect of making the contours more spherical, in the FOG regime from less-than-unity satellite velocity bias and in the Kaiser regime from central velocity bias. } \label{fig:xipno}
\end{figure}

The contribution to the bump in the quadrupole $s^2\xi_2$ at $s\sim$3--5$\mpchi$ mostly arises from the one-halo central-satellite and two-halo central-central galaxy pairs (Zheng \& Guo in preparation), with the former related to the FOG effect and the latter to the Kaiser effect. Since the quadrupole from the FOG effect is positive and that from the Kaiser effect is negative, it is easier to see the velocity bias effects in the 3D $\xi(r_p,r_{\rm\pi})$, as illustrated in Fig.~\ref{fig:xipno} for the best-fitting HOD model.

The case of removing the velocity bias (by setting $\alpha_c=0$ and $\alpha_s=1$) is shown in Fig.~\ref{fig:xipno} as black contours. The red (blue) contours correspond to the case of only allowing for satellite (central) velocity bias. A less-than-unity satellite velocity bias parameter means that satellite galaxies are cooler, and in redshift space the central-satellite pairs are less stretched along the LOS. Therefore, the effect is to make the FOG feature significantly less elongated. The central galaxy bias has a weaker effect on the FOG, because the redshift-space displacement of the central-satellite pairs is governed by the quadrature sum of $\alpha_c$ and $\alpha_s$ and is dominated by $\alpha_s$ for $\alpha_c\sim 0$ and $\alpha_s\sim 1$. The central velocity bias does have a strong effect on the Kaiser feature. A non-zero central velocity bias smears the two-halo central-central pairs along the LOS (i.e., smoothing the LOS pair separation), which tends to reduce the Kaiser effect. As seen in Fig.~\ref{fig:xipno}, the overall effect of a low satellite velocity bias $\alpha_s$ is to make the contours more squashed along the LOS direction, while a high central velocity bias $\alpha_c$ has the opposite effect. Therefore, to some degree, the two effects can compensate each other, which gives rise to the (weak) anti-correlation between $\alpha_c$ and $\alpha_s$ seen in Fig.~\ref{fig:alphacs}.

Another way of summarizing the velocity bias impact is that a less-than-unity satellite velocity bias makes the contours in the FOG regime more spherical, and a non-zero central velocity bias make the contours in the Kaiser regime more spherical. A comparison between the $\xi(r_p,r_{\pi})$ contours for the case without velocity bias in Fig.~\ref{fig:xipno} and the measurement in Fig.~\ref{fig:xip} shows that the contours in the measurements appear to be more spherical in both the FOG and Kaiser regime than the case without velocity bias. For the model to match the data, a less-than-unity satellite velocity bias is needed to fit the FOG feature, and a non-zero central velocity bias is required to make the Kaiser part more spherical. The overall effect leads to an increase in $\xi_2$, compared to the case without velocity bias (as shown in Fig.~\ref{fig:wpxi}).

In Fig.~\ref{fig:wpxi} the deviation of the dotted curve (case without velocity bias) from the measurements of the quadrupole appears stronger for the FOF haloes. This behaviour is consistent with a lower value of the satellite velocity bias parameter inferred in the FOF model than in the SO model. The FOF algorithm has the possibility of linking two distinct haloes through some of the interhalo particles. Such a bridging effect makes the satellite distributions (therefore the separations of central--satellite pairs) in the bridged haloes more spread along the LOS in redshift-space, enhancing the FOG effect. To reduce the effect to match the measurements, a lower value of the satellite velocity bias is required.

With velocity bias parameters fixed at $\alpha_c=0$ and $\alpha_s=1$, we fit the $w_p$ and multipoles by varying the other HOD parameters related to the occupation function, i.e., using the commonly adopted five-parameter HOD model without velocity bias. The results are shown as the solid curves in Fig.~\ref{fig:wpxi}. The best-fitting $\chi^2$ values with this `no-velocity bias' case for the SO and FOF models are $79.2$ and $142.2$, respectively. As expected, these best-fitting $\chi^2$ values are reduced compared to the dotted-lines case, but the models still do not provide a good fit to the measurements. The main discrepant feature is again the quadrupole on scales of a few Mpc. Clearly, the data require the inclusion of velocity bias.

We can determine what halo masses our velocity bias constraints are most sensitive. We start from the best-fitting model and turn off the central or satellite velocity bias for galaxies in haloes above a given mass threshold. The value of $\chi^2$ increases substantially in the mass range of $\log M=$13--13.6 for central velocity bias and $\log M=$13.6--14 for satellite velocity bias. As an estimate, we adopt $\log M=13.3$ ($\log M=$13.8) as the representative mass of haloes for which the central (satellite) velocity bias is most sensitive, corresponding to the peak of the distribution of host halo mass for central (satellite) galaxies \citep{Guo14}. At $z\sim 0.5$, the 1D velocity dispersion of dark matter in $\log M=13.3$ and $13.8$ haloes is ${\sim}315$ and $462\,{\rm km\, s^{-1}}$, respectively. Therefore, for the SO model, the best-fitting central velocity bias ($\alpha_c=0.22$) corresponds to a velocity dispersion of $69\,{\rm km\, s^{-1}}$, and that for satellites ($\alpha_s=0.86$) is $397\,{\rm km\, s^{-1}}$.

Finally, we emphasize that the central velocity bias is defined with respect to the halo core, defined by the inner 25 per cent of particles around the potential minimum. Our result shows that central galaxies have an additional motion
relative to the halo core. The result can also be cast into the relative motion between the galaxy and the whole halo (i.e., the halo centre-of-mass velocity). From the simulation, we find that the velocity dispersion between the halo core and centre of mass is $\sigma_{\rm core-cm}=\alpha_{\rm core-cm}\sigma_v\sim 0.158\sigma_v$. The conversion to velocity bias parameters $\tilde{\alpha}_c$ and $\tilde{\alpha}_s$ in the centre-of-mass frame can be done by noticing that $\tilde{\alpha}_c^2\sigma_v^2=\alpha_c^2\sigma_v^2+\sigma_{\rm core-cm}^2$ and $\tilde{\alpha}_s^2\sigma_v^2=\alpha_s^2\sigma_{\rm p-core}^2+\sigma_{\rm core-cm}^2$, where $\sigma_{\rm p-core}$ is the velocity dispersion of particles in the frame of the core and it takes the value
$\sigma_{\rm p-core}^2=(1-\alpha_{\rm core-cm}^2)\sigma_v^2$. That is, we have $\tilde{\alpha}_c^2=\alpha_c^2+\alpha_{\rm core-cm}^2$ and $\tilde{\alpha}_s^2=\alpha_s^2(1-\alpha_{\rm core-cm}^2)+\alpha_{\rm core-cm}^2$. Plugging in the numbers, we infer that with respect to the centre of mass, the central and satellite velocity bias
parameters are $0.27_{-0.03}^{+0.03}$ and $0.86_{-0.03}^{+0.08}$ (for the SO model) and $0.28_{-0.02}^{+0.02}$ and $0.70_{-0.04}^{+0.03}$ (for the FOF model), respectively.

\subsection{Tests of the modelling results}
\label{sec:sys} In this subsection, we perform several investigations to better understand the nature of the velocity bias constraints and to test the robustness of the results. We first consider the effects from changing the data, including using different combinations of the clustering measurements, removing the small-scale data points, using a fibre-collision free sample, and dependence of the velocity bias on galaxy luminosity. We then examine variations in the model, focusing on different changes in the HOD parametrization and the effect of varying the spatial distribution of satellites inside haloes. Finally, we present a brief discussion on the dependence on cosmology.

In view of the bridging effect for FOF haloes, we regard the results from the SO model as more reliable, and hereafter focus our efforts on models based on SO haloes. We wish to emphasize, however, that in both the SO and FOF models we define the halo centre as the position of potential minimum. The consistency between the central velocity bias constraints from the two models (Fig.~\ref{fig:alphacs}) implies that as long as the positions of potential minimum are adopted as centres, the results on the central velocity bias are insensitive to the exact definition of haloes.

\begin{figure}
\includegraphics[width=0.38\textwidth]{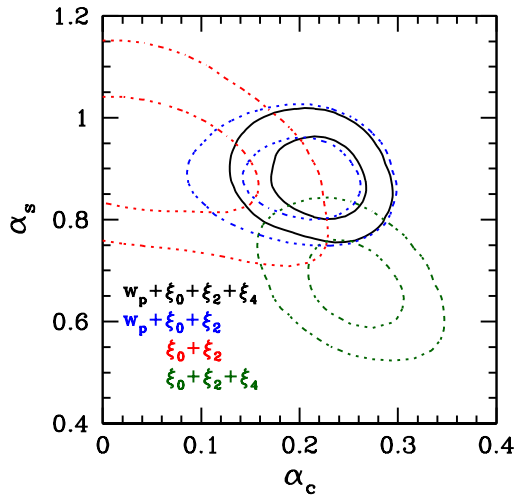}\caption{
Velocity bias constraints from modelling different combinations of the four measurements of $w_p$, $\xi_0$, $\xi_2$ and $\xi_4$. The results from our fiducial model are shown as the solid contours, while those from other combinations are shown as the dotted contours of different colours.} \label{fig:alphacs_test}
\end{figure}
{\it Constraints from various combinations of the measurements.} Our main results in the previous subsection are based upon jointly modelling the projected 2PCF and three redshift-space 2PCF multipole moments, $w_p$, $\xi_0$, $\xi_2$, and $\xi_4$. We now explore the constraints on velocity bias using different combinations of the measurements, shown in Fig.~\ref{fig:alphacs_test} as contours with different colours. Using only the monopole and quadrupole ($\xi_0+\xi_2$) leads to weaker constraints on the galaxy velocity bias. Adding the hexadecapole ($\xi_0+\xi_2+\xi_4$) provides much tighter constraints, consistent with the recent result of \cite{Hikage14}. The reason for the improvement is that the hexadecapole $\xi_4$ is more sensitive to small angular variations in the 2PCF, picking up finer features in the FOG than $\xi_0$ and $\xi_2$ alone. Including $w_p$ provides even stronger constraints on the velocity bias (the $w_p+\xi_0+\xi_2$ and $w_p+\xi_0+\xi_2+\xi_4$ cases). On linear scales, the redshift-space 2PCF (hence $w_p$) is fully determined by $\xi_0$, $\xi_2$, and $\xi_4$ \citep{Kaiser87,Hamilton92}. However, on nonlinear scales, especially in the FOG regime, higher-order multipoles ($l>4$) exist. The small-scale $w_p$ therefore encodes some additional information about the higher-order moments. The inclusion of $w_p$ tightens the constraints on the velocity bias, mainly by excluding certain regions in the parameter space spanned by the parameters in the mean occupation functions. The best-fitting $\alpha_s$ from fitting $\xi_0+\xi_2+\xi_4$ is significantly smaller than the one with $w_p$ included, because the small-scale shape of $w_p$ is still sensitive to the satellite velocity bias, especially for CMASS galaxies in these massive haloes.

\begin{figure}
\includegraphics[width=0.38\textwidth]{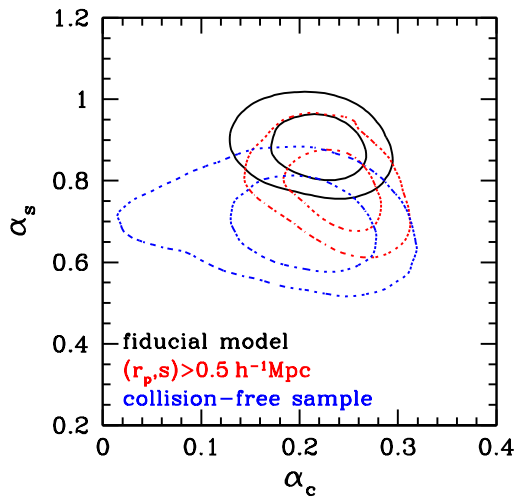}
\caption{Velocity bias constraints when removing the small-scale($<0.5\mpchi$) measurements (dotted red contours) and when using the collision-free sample (dotted blue contours), compared to those from the fiducial model (solid black contours).} \label{fig:alphacs_dropsmall}
\end{figure}
\begin{figure}
\includegraphics[width=0.38\textwidth]{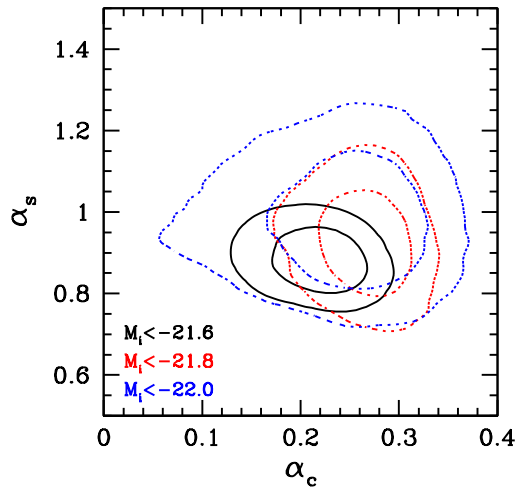}
\caption{Luminosity-dependence of the velocity bias. Velocity bias constraints from galaxy samples with different luminosity thresholds, $M_i<-21.6$ (fiducial model), $M_i<21.8$, and $M_i<-22.0$, all at $0.48<z<0.55$.} \label{fig:alphacs_lum}
\end{figure}
{\it Constraints from removing small-scale data points.} In our main analysis, the smallest scale measured is $0.13\mpchi$ (in $r_p$ or $s$). At $z\sim 0.5$, the fibre-collision scale (the physical separation corresponding to the angular fibre collision constraint) is about $0.4\mpchi$. We apply the method developed by \citet{Guo12} to account for the fibre collisions in both projected and redshift spaces. This technique is shown to be accurate and unbiased in \citet{Guo12} and in Appendix~\ref{app:fiber} of this paper. Nevertheless, to ensure that the results are not affected by any remaining small-scale systematics in the fibre-collision correction and to obtain a sense of the constraining power of the small-scale measurements, we repeat our analysis when removing the data points with $r_p<0.5\mpchi$ and $s<0.5\mpchi$. Fig.~\ref{fig:alphacs_dropsmall} shows the constraints when keeping only the  $>0.5\mpchi$ data points. The constraint on the central velocity bias is not significantly affected, since it is produced mainly from scales larger than $\sim 1\mpchi$ (contributed mostly by two-halo central-central pairs). The satellite velocity bias is slightly less constrained and shifts towards somewhat lower values, reflecting the fact that it arises mainly from the FOG regime (through the contribution from one-halo central-satellite pairs). Overall, removing the small-scale points does not lead to any substantial change in the velocity bias constraints.

{\it Constraints from a sample constructed by complete sectors.} As a further check for any possible effect of the fibre collision correction, we model the clustering measurement for a sample of galaxies free of fibre collision. We construct the sample in the BOSS DR11 from the (non-contiguous) set of all complete sectors, where there are no collided galaxies. The same selection criteria (e.g., luminosity threshold and redshift range) as our main sample are applied. The total effective area (and hence volume) of the complete sectors sample is about a quarter of that of our fiducial sample. The modelling result for this fibre-collision free sample from complete sectors is shown as the dotted blue contours in Fig.~\ref{fig:alphacs_dropsmall}. Although the uncertainties become larger, the velocity bias constraints are still consistent with those from the main sample. The test implies that our fibre-collision correction method works well, and our constraints on velocity bias are not an artefact of inaccurate fibre-collision corrections.

{\it Dependence of the velocity bias on galaxy luminosity.} In this paper, we focus our modelling effort on the volume-limited, luminosity-threshold sample with $M_i<-21.6$. We also construct volume-limited sample with higher luminosity thresholds, $M_i<-21.8$ and $M_i<-22.0$, in the same redshift range of $0.48<z<0.55$ \citep{Guo14}. The modelling results for these brighter, but lower number density, galaxy samples are shown in Fig.~\ref{fig:alphacs_lum}. The constraints on velocity bias are not as strong as the ones from the $M_i<-21.6$ sample, but are generally consistent with them. In our velocity bias parametrization, we do not introduce any dependence on halo mass. From the above test, no strong dependence of the velocity bias on galaxy luminosity is found. Therefore, with the CMASS sample, in the luminosity range considered, we do not expect a strong dependence of the velocity bias on halo mass, and our choice of the velocity bias parametrization appears to be justified. The test also demonstrates that the velocity bias inferred from the $M_i<-21.6$ sample is unlikely caused by some peculiarity in the sample (e.g., from sample variance).

\begin{figure}
\includegraphics[width=0.38\textwidth]{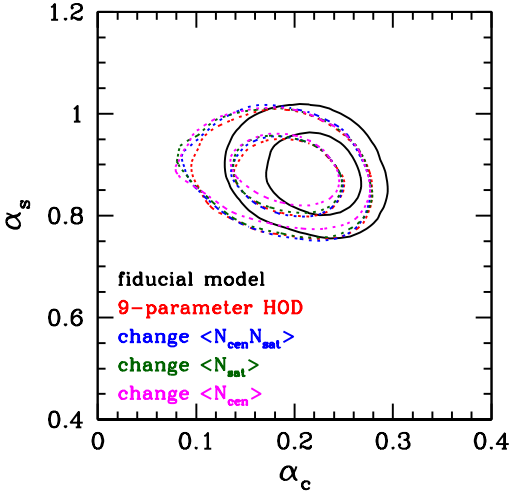}
\caption{Velocity bias constraints with different variations or more flexibilities of the HOD model (see text for details). The constraints resulting from the various HOD models are shown as the dotted contours with different colors, while those from the fiducial model are represented by the solid contours. } \label{fig:alphacs_par}
\end{figure}

{\it Constraints from variations in the HOD parametrization.} In our HOD parametrization, we adopt the commonly used five-parameter form for the mean occupation functions [see equations~(\ref{eqn:Ncen}) and (\ref{eqn:Nsat})]. While such a parametrization is sufficient to describe a luminosity-threshold sample \citep{Zheng05,Zheng07,Zehavi11}, to test whether the velocity bias constraints are affected by a too-restrictive form of the mean occupation functions, we also perform modelling by varying the HOD parametrization.

For the central velocity bias, one may wonder whether it is in fact caused by the motion of satellites in haloes whose central galaxies fall out of the sample selection. The situation could occur if the central galaxy is not the brightest galaxy in a halo and its luminosity falls short of the threshold for the sample, or if the photometric error causes the central galaxy scatter out of the sample \citep[e.g.,][]{Reid14}. It is possible that in our sample, there might be some satellite galaxies that are the single occupant of their parent haloes. To test this scenario, we  modify the correlation between central and satellite galaxies and the mean occupation functions, and present the results in Fig.~\ref{fig:alphacs_par}. In our fiducial model, when computing one-halo central-satellite pairs, we assume that no satellite can be brighter than the central galaxy in a halo, i.e., satellites can only occupy haloes where there are central galaxies. We first relax the assumption of the dependence between central and satellite occupations by setting $\langle N_{\rm cen} N_{\rm sat} \rangle = \langle N_{\rm cen} \rangle \langle N_{\rm sat} \rangle$ \citep[e.g.,][]{Zheng05,Guo14}. This approach leads to only a slight shift in the central velocity bias (blue dotted contours in Fig.~\ref{fig:alphacs_par}). We further remove the modification to the satellite occupation function from the central satellite occupation profile in equation~(\ref{eqn:Nsat}), and the result is similar to the above case (green dotted contours).

To explicitly address the possibility that some haloes do not have central galaxies satisfying the sample selection, we consider a case where $\langle N_{\rm cen}\rangle$ at the high mass end approaches $f_a$, instead of unity (i.e., the right-hand side of equation~(\ref{eqn:Ncen}) is multiplied by $f_a$). The best-fitting model has $f_a=0.46$ with $\chi^2=44$, i.e. the model can even accommodate losing about half of the central galaxies; However, the constraints on the central and satellite velocity bias change only slightly (dotted magenta contours in Fig.~\ref{fig:alphacs_par}).

Finally, we introduce a more flexible HOD parametrization. The free parameter $n$ is added in the mean central occupation function, changing the curvature of $\langle N_{\rm cen}(M)\rangle$ in equation~(\ref{eqn:Ncen}) by using the form $\{[1+{\rm erf}()]/2\}^n$. We further add the free parameter ${\rm d}\alpha/{\rm d}\log M$ to the mean satellite occupation function, i.e., we introduce a running in the power-law index of $N_{\rm sat}(M)$ in equation~(\ref{eqn:Nsat}). With this nine-parameter model (seven for the mean occupation functions and two for velocity bias), the constraints on the two velocity bias parameters again only slightly change, as shown by the dotted-red contours in Fig.~\ref{fig:alphacs_par}.

The tests show that satellite galaxies in haloes with no central galaxies can not play the role of producing the detected central velocity bias. This result is easy to understand -- the satellite fraction of the sample is small, only about 7 per cent overall. Limited to the halo mass range that the central velocity bias is sensitive to ($\log M=$13--13.6; see \S~\ref{sec:main_results}), the contribution from satellites is almost negligible. The tests also demonstrate that our fiducial HOD parametrization is not too restrictive. Adding additional flexibility to the parametrization does not significantly change the results. The existence of the velocity bias is therefore not an artifact caused by the assumed HOD parametrization.

\begin{figure*}
\includegraphics[width=0.7\textwidth]{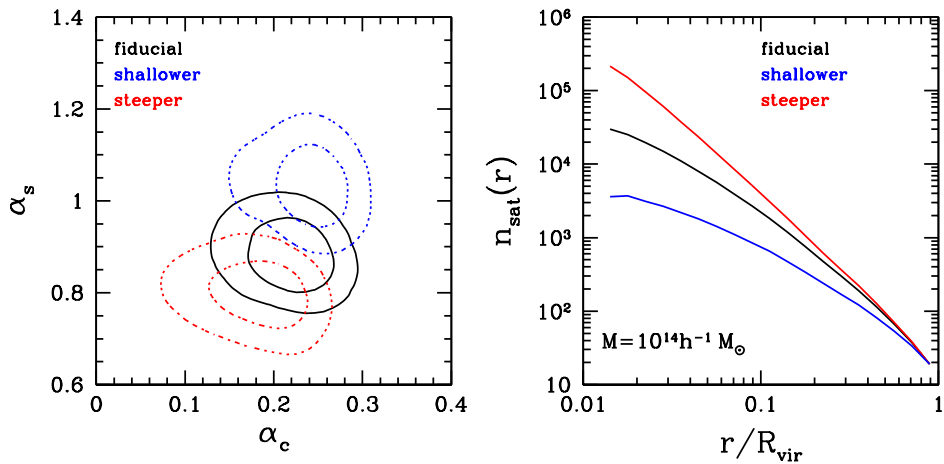}\caption{
Velocity bias constraints for steeper and shallower halo profiles of satellite galaxies (see the text). Left-hand panel: comparison of the constraints with different satellites spatial distribution profiles. Right-hand panel: an example of the different satellite spatial distribution profiles inside haloes of ${\sim}10^{14}\msun$. The displayed profiles are normalized to have the same value at the virial radius $R_{\rm vir}$ for comparison. } \label{fig:alphacs_pro}
\end{figure*}
{\it Effect of varying the spatial distribution of satellites inside haloes.} In our fiducial model, we use the positions of random dark matter particles inside haloes to represent satellite galaxies. We then modify the particle velocities according to the value of satellite velocity bias parameter $\alpha_s$. In principle, for a steady state system, a change in the velocity distribution is followed by a change in the spatial distribution \citep[e.g.,][]{Bosch05}. Building a fully self-consistent model is computationally expensive. Here, instead, we investigate the effect of changes in the satellites spatial distribution on the velocity bias obtained. We consider two extreme cases of the satellite number density distribution profile $n_{\rm sat}(r)$: one much shallower than the dark matter distribution and the other much steeper,
%%%%%%%%%%%%%%%%%%%%%%%%%%%%%%%%%%%%%%%%%%%%%%%%%%
\begin{eqnarray}
\rm{shallower:}&\quad&n_{\rm sat}(r)\propto\rho_m(r)(r/R_{\rm vir})^{0.5}, \\
\rm{steeper:}&\quad&n_{\rm sat}(r)\propto\rho_m(r)(1+0.1R_{\rm vir}/r),
\end{eqnarray}
%%%%%%%%%%%%%%%%%%%%%%%%%%%%%%%%%%%%%%%%%%%%%%%%%%
where $\rho_m(r)$ is the dark matter distribution profile, and $R_{\rm vir}$ is the virial radius of the dark matter halo. Satellites are assigned to the dark matter particles according to the corresponding ratios of $n_{\rm sat}(r)/\rho_m(r)$. The shallower and steeper profiles, together with the fiducial one, are shown in the right-hand panel of Fig.~\ref{fig:alphacs_pro} for haloes of $\sim 10^{14}\msun$, and are normalized to have the same value at $R_{\rm vir}$.

The resulting model constraints on $\alpha_c$ and $\alpha_s$ are presented in the left-hand panel of Fig.~\ref{fig:alphacs_pro}. The shallower satellite distribution profile leads to a significantly larger $\alpha_s$, while the steeper profile leads to a somewhat smaller value of $\alpha_s$. This behaviour is expected --- lower velocity dispersions are needed to compensate the enhancement in the small-scale FOG effect caused by a steeper satellite profile, and vice versa. The constraint on the central velocity bias $\alpha_c$, however, only shows a small increase, even with the extreme changes in the satellite profile. The best-fitting $\chi^2$ for the steeper and shallower profiles are $69.5$ and $72.7$, respectively, larger than that of our fiducial model (55.9). Furthermore, realistic galaxy distributions are unlikely to have such extreme profiles. The dark matter density profile in our fiducial model can be well characterized by the Navarro-Frenk-White (NFW) profile \citep{Navarro97}. The NFW profile does provide a reasonably good description of the satellite galaxy distribution for CMASS galaxies \citep{Guo14}. A stacking analysis with SDSS galaxy groups and clusters shows that massive galaxies may have slightly shallower profiles than the NFW profile \citep{Wang14}, but the largest deviation at $r\sim0.1R_{\rm vir}$ is only ${\sim}13$ per cent, much smaller than those from our two extreme cases. Thus, we expect that more realistic changes in the satellite distribution profile will lead to $\alpha_c$ and $\alpha_s$ constraints similar to these of our fiducial model (the solid contours in Fig.~\ref{fig:alphacs_pro}).

For a model with self-consistent satellite velocity and spatial distributions, we would, in general, obtain radius-dependent velocity bias (under simplifying assumptions such as steady state, generalized NFW profile, and spherical symmetry; e.g., \citealt{Bosch05}). We would then obtain self-consistent velocity bias contours (with radially averaged satellite velocity bias). Even in the absence of such a model, our results demonstrate that satellite velocity bias is needed if we assume that they follow the spatial distribution of dark matter inside haloes. Or, conversely, if there is no satellite velocity bias, their spatial distribution must differ from that of matter. This result implies that the velocity and spatial distributions of satellites cannot {\it both} be unbiased (which is a $1.9\sigma$ result from our analysis), which may be a more appropriate statement of the results of the satellite velocity bias. On the other hand, our test shows that the central velocity bias is robust, and cannot be mitigated by varying the distribution of satellites.

{\it Dependence on cosmology.} Our model is based on the MultiDark simulation, which has $(\Omega_m, \sigma_8)=(0.27, 0.82)$. These cosmological parameters are consistent with the values determined by combinations of various data sets \citep[e.g.,][]{Anderson13}. Nonetheless, it is important to consider whether the inferred velocity bias depends on the cosmology assumed (i.e. whether a more accurate cosmological model, within the range allowed by current constraints, can mitigate our detection of velocity bias).  It would also be useful to know how the velocity bias constraints depend on cosmology, or how the small-scale redshift clustering may tighten cosmological constraints \citep[e.g.,][]{Tinker06,Tinker07}.  We present some simplified estimates of this here.

The velocity bias impacts the quadrupole on scales of a few Mpc (see Fig.~\ref{fig:wpxi}) and the central velocity bias is mainly for haloes of a few times $10^{13}\msun$ (see Section~\ref{sec:main_results}). For a simple estimate, we consider the two-halo central-central term of the redshift 2PCF on scales of $\sim 5\mpchi$. If the central velocity bias effect in our fiducial cosmological model (${\rm Cosmology_{fid}}$) were to be fully accounted for by another cosmological model (${\rm Cosmology_{new}}$), it would require the 1D halo-halo velocity dispersion $\sigma_{hh}$ in the new cosmology to absorb the joint effect of halo-halo velocity dispersion and the velocity dispersion of central galaxies ($\alpha_c\sigma_{1D}$ with $\sigma_{1D}$ the velocity dispersion inside haloes) in the fiducial model. In this case, we would have $\sigma^2_{hh}(M|{\rm Cosmology_{new}})=\sigma^2_{hh}(M|{\rm Cosmology_{fid}}) +\alpha_c^2\sigma^2_{1D}(M|{\rm Cosmology_{fid}})$ at $z\sim 0.5$. With the appropriate scaling of $\sigma_{hh}$ and $\sigma_{1D}$ with cosmology and halo mass \citep[e.g.,][]{Zheng02,Tinker06}, we find that the required change in $\Omega_m$ would be substantial, from 0.27 to 0.345. Our estimate is based on a single mass and a single scale, but nonetheless provides a sense of the required change. For all the measurements with spatial clustering encoding dynamical information, it is unlikely that the HOD (including the velocity bias) and cosmology will be degenerate at this level \citep{Zheng07b}. A preliminary test with a cosmological model with $\Omega_m=0.315$ (consistent with the constraints from Planck; \citealt{PlanckCollaboration13}) shows that the best-fitting $\alpha_c$ and $\alpha_s$ values decrease by only about 10 per cent. We conclude that our velocity bias constraints are robust against reasonable changes in cosmology.

The various tests presented in this subsection suggest that our inference of the galaxy velocity bias parameters is robust, especially for the central galaxies.

\section{Discussion and Implications}
\label{sec:implications}

By modelling the projected 2PCF and the redshift-space 2PCF multipoles of a well-defined, volume-limited sample of CMASS BOSS galaxies, we find that massive central galaxies are not at rest with respect to the halo core (defined by the inner 25 per cent of particles around the position of potential minimum), while massive satellite galaxies appear to be ``cooler'' than the dark matter particles inside haloes. Note that the central velocity bias is defined with respect to the halo core, which can also be cast relative to the halo centre of mass.

What could lead to such a difference in the velocity distribution between galaxies and dark matter?
The difference clearly implies that the galaxies and their host halo are not mutually relaxed \citep[e.g.][]{Bosch05}. Haloes evolve by continuously accreting matter or merging, and naturally we expect that the outer part of the halo is not well relaxed. In the inner (core) region, however, the halo is not completely relaxed either, as implied by the fact that the mean velocity of dark matter particles around the centre depends on radius \citep[e.g.][]{Behroozi13,Reid14}. Compared to dark matter particles, galaxies experience additional processes, like dynamical friction, tidal disruption, and mergers. Dynamical friction can slow down the motion of satellite galaxies inside haloes, and the effect is stronger for more massive satellites \citep{Chandra42,BT08}. Satellites with eccentric orbits are more likely to be tidally stripped or disrupted during the peri-centre passage. The central galaxy in a halo can be disturbed by a satellite merger. All these processes can contribute to differences in the dark matter and galaxy velocity distributions.

\citet{Yoshikawa03} study the velocity distribution of central and satellite galaxies, based on a smoothed particle hydrodynamics (SPH) simulation. Converting the velocity distributions to velocity bias parameters consistent with our definitions (e.g., velocity dispersion ratio between central/satellite galaxies and dark matter), the results in their Figure 10 imply $\alpha_c\sim 0.3$ and $\alpha_s\sim 0.8$ (for the brightest satellite). In the halo mass range to which our inferred satellite velocity bias is sensitive, we expect on average $\lesssim 1$ satellite in a halo, so our satellite velocity bias is effectively determined by the most luminous satellites. \citet{Berlind03} also investigate the velocity distributions with a SPH galaxy formation simulation; they found $\alpha_c\sim 0.2$ and $\alpha_s\sim 0.9$ (but for all satellites, not the most luminous ones) in haloes of mass $\log M=$13--14. \citet{Wu13b} investigate velocity bias of satellite galaxies in galaxy clusters using $N$-body and hydrodynamical simulations. In the $N$-body simulations, subhaloes are rank-ordered according to circular velocity to serve as galaxy tracers. They report that the most luminous satellites are cooler than dark matter particles, and the results do not depend strongly on redshift and halo mass. They also found that the satellite velocity bias depends on galaxy mass or luminosity, which can significantly impact results based upon using satellite kinematics for cluster mass estimates \citep[e.g.,][]{Goto05,Old13}. A simple extrapolation of the results in Figures 1 and 2 of \citet{Wu13b} yields a satellite velocity bias of around 0.7 for the most luminous satellites. All the above conclusions should be compared to our FOF results (cast relative to the centre of mass) to be consistent with the halo and velocity definitions in these papers. Overall, the values of velocity bias of central/satellite galaxies we infer from galaxy clustering in this work are at the level expected by galaxy formation theory.

\citet{Bosch05} investigated the motion of central galaxies with respect to the mean motion of satellites by comparing galaxy groups in the Two-Degree Field Galaxy Redshift Survey \citep{Colless99} and in the SDSS with those in mock catalogues. Note that the mean motion of satellites is a proxy of the halo centre-of-mass velocity (modulated by the satellite spatial and velocity bias). They infer a central velocity bias in the range of 0.2--0.6 for galaxies in massive groups
and clusters (not corrected for possible satellite velocity bias). A similar analysis by \citet{Skibba11} concludes that a central velocity bias $\lesssim 0.2$ is consistent with the data of SDSS galaxy groups. Our inferred central velocity bias (relative to the centre of mass) is broadly consistent with the values derived from group catalogues. Using the mean motion of satellites as a proxy of the halo centre-of-mass velocity, \cite{Lauer14} find a central velocity bias for brightest cluster galaxies in Abell clusters to be around 0.55. This is much larger than our value, but we note that the haloes (above $10^{14}\msun$) are much more massive than the ones relevant to our results ($\sim 2\times 10^{13}\msun$).

The existence of velocity bias of central galaxies means that they are not at rest with respect to halo cores defined by 25 per cent of inner particles around the position of halo potential minimum. An immediate implication is that the central galaxy is not always at the centre of its host halo core, but rather it oscillates around the centre, leading to a spatial offset from the centre. The offset $r_{\rm cen}$ can be roughly estimated by assuming that the central galaxy moves on a circular orbit at this radius with velocity $v_c$ in an NFW halo potential. For a set of randomly oriented circular orbits, the 1D velocity dispersion is $\sigma_c=v_c(r_{\rm cen})/\sqrt{3}$. To a good approximation the 1D velocity dispersion of dark matter in a halo with NFW profile is $\sigma_v=v_c(R_{\rm vir})/\sqrt{2}$. By setting $\sigma_c=\alpha_c\sigma_v$, we find that $r_{\rm cen}\simeq 3[\ln(1+c)/c-1/(1+c)]\alpha_c^2 r_s$, where $r_s=R_{\rm vir}/c$ is the scale radius and $c$ the concentration parameter of the dark matter halo. For $\alpha_c\sim 0.2$ and the mass range relevant to our constraints ($\log M=$13--13.6), we find that $r_{\rm cen}$ is about 2.2 per cent of $r_s$ or 0.4 per cent of $R_{\rm vir}$, which is quite close to that found from assuming a parametrized distribution of central galaxy positions \citep[e.g.,][]{Bosch05}. This offset translates to a mean projected radius of $\sim 0.3$ per cent of $R_{\rm vir}$, or $\sim$1--3$h^{-1}{\rm kpc}$ in the above mass range. This offset is much smaller than the smallest scale of our data points, and there is virtually no effect on the results by completely neglecting it in our model. It is also clear that the effect of off-centre central galaxies is much easier to detect in velocity space rather than in real space \citep{Bosch03}.

In the above scenario, the motion of the central galaxy may just reflect that of the potential minimum (with the core itself not relaxed). However, given the expected different evolution histories of dark matter and baryons, it is likely that the central galaxy is not at rest with respect to the potential minimum position, and we expect that the two have an offset at a level similar to our above estimate. Such an offset between central galaxies and halo potential minimum (or most bound dark matter particles) at the above level is indeed seen in SPH galaxy formation simulations \citep[e.g.,][]{Berlind03}. Observationally, detecting such an offset is not straightforward, because we need an appropriate way to probe the potential minimum. Offsets between X-ray or Sunyaev-Zel'dovich (SZ) effect peaks/centroids and the brightest central galaxies (BCGs) in galaxy clusters are observed, usually at the level of 10 kpc or larger \citep[e.g.,][]{Ascaso11,Mann12,Song12,Linden14}. However, hot gas distribution from X-ray and SZ effect does not necessarily follow the gravitational potential closely given its collisional nature (with, e.g., the bullet cluster as an extreme example; \citealt{Clowe06}). Density peaks or centroids from gravitational lensing may be a better probe of the potential. Offsets between X-ray and lensing centres are in fact observed at a level of tens of kpc \citep[e.g.,][]{Allen98,Shan10,George12}. \cite{George12} infer an offset around $20 h^{-1} {\rm kpc}$ between BCGs and lensing centres. Based on a strong lensing mass-modelling technique, \citet{Zitrin12} characterize the offset between BCGs and dark matter projected centre in SDSS clusters and infer a median offset about 30$h^{-1}{\rm kpc}$ for $z\sim 0.5$ clusters. The offsets are larger than our simple estimate based on velocity bias. Although our estimate is not sophisticated, it is more likely that the difference reflects the difference between lensing peaks and lensing centroid and our lower halo mass scales.

The motion of central galaxies relative to the halo cores and the likely offsets have many important implications. The effect should be taken into account in lensing studies of galaxy groups and clusters for accurately mapping out the dark matter density profile \citep[e.g.,][]{George12}. The offset between the central galaxy and the core mass distribution can contribute to the external shear needed in modelling strong lensing systems \citep[e.g.,][]{Keeton97,Wong11, Gavazzi12}. Their relative motion and offset can cause continuous tidal interactions, which might heat the galaxy, cause instabilities, and disturb the galaxy's morphology \cite[e.g.,][]{Bosch05}. If we boldly extrapolate the likely offset between central galaxy and potential minimum to Milky Way and lower-mass haloes, such an effect needs to be considered when searching for dark matter annihilation signals \citep[see, e.g.,][for such an offset found in simulations of Milky Way-like galaxies]{Kuhlen13}.

Finally, the velocity bias can also impact the use of redshift-space clustering to constrain cosmology and gravity. In particular, redshift-space distortions are a sensitive probe of $f\sigma_8$ \citep[e.g.,][]{Guzzo08, Percival09,Reid12,Reid14,Samushia13}, where $f={\rm d}\ln D/{\rm d}\ln a$ is the growth rate (the fractional change of the growth factor $D$ with respect to that of the scale factor $a$). Neglecting the central and satellite velocity bias in modelling the redshift-space 2PCFs, especially when data points in the trans-linear or nonlinear scales are included, can introduce potential systematic errors in the $f\sigma_8$ constraints \citep[e.g.,][]{Wu13a}. The commonly adopted FOG model with different forms of damping functions may partly account for the velocity bias effect, but residual systematic errors at the per cent level are still expected \citep[][]{Torre12}, which will be important for pristine measurements with future large surveys.

\section{Conclusion}
\label{sec:conclusions}

In this paper, we measure and perform HOD modelling of the small- to intermediate-scale projected 2PCF and redshift-space 2PCF for a volume-limited, luminosity-threshold sample of $z\sim 0.5$ CMASS galaxies in SDSS-III BOSS. Built on the MultiDark simulation, our model is equivalent to populating haloes in the simulation with galaxies. Besides the usual HOD parameters for the mean occupation function, we also introduce two additional velocity bias parameters to characterize the galaxy velocity distribution relative to that of the dark matter. The redshift-space clustering data require differences between the velocity distributions of galaxies and dark matter inside haloes, and the constraints arise mainly from the one-halo FOG feature and small-scale two-halo clustering.

Our joint fitting of the projected and redshift space 2PCFs reveals that the central galaxies (in haloes of a few times $10^{13}\msun$ haloes) move with respect to the halo cores (defined as the inner 25 per cent particles around the halo potential minimum; see Appendix~\ref{app:full} for a discussion on the dependence of the velocity bias on the definition of halo cores).
It implies that the central galaxy and the halo core are not
mutually relaxed. In the rest frame of the core, the velocity dispersion of central galaxies is $\alpha_{\rm c}=0.22^{+0.03}_{-0.04}$ times that of dark matter particles inside haloes (with the SO halo model), i.e. the specific kinetic energy of central galaxies is about 4.5 per cent that of dark matter particles. Since
halo cores are not at rest with respect to the centre of mass of haloes, this
becomes $\alpha_{\rm c}=0.27^{+0.03}_{-0.03}$ when cast in the
centre-of-mass frame of haloes. The motion implies a typical offset between the central galaxy and the core centre at the level of $\lesssim$1 per cent of the virial radius. The results appear robust to numerous consistency checks.

The satellite galaxies (roughly the most luminous ones in $\sim 10^{14}\msun$ haloes) are cooler than the dark matter particles, with velocity dispersions about $\alpha_{\rm s}=0.86^{+0.08}_{-0.03}$ times those of the dark matter particles, or $\sim$79 per cent in terms of specific kinetic energy. Given the connection between kinematics and spatial distribution, a more appropriate statement of the satellite velocity bias results is that the velocity and spatial distributions of satellite galaxies cannot both be unbiased (at a level of $1.9\sigma$). We do not find evidence for strong luminosity dependence of the velocity bias parameters.

The measured galaxy velocity bias is in broad agreement with predictions from galaxy formation simulations. The existence of galaxy velocity bias is related to the evolution of galaxies inside haloes, encoding information like dynamical friction, tidal stripping and disruption, and mergers. It has important implications for a variety of applications related to galaxy formation and cosmology. Redshift-space higher-order clustering statistics, e.g. the three-point correlation functions, can help tighten the constraints on galaxy velocity bias \cite{Guo14b}. The investigation in this paper focuses on $z\sim 0.5$ massive galaxies. We plan to extend the work to galaxies of different types at different redshifts, to study the dependence of velocity bias on the properties of galaxies and its evolution. This will provide more insight about its origin and potential implications.

\section*{Acknowledgements}
We thank Y.~P. Jing, Alexie Leauthaud, Cheng Li, Surhud More, Beth Reid, Uros Seljak, and Hee-Jong Seo for helpful discussions. We thank the anonymous referee for the helpful comments. We thank Kristin Riebe for the help in obtaining the MultiDark Simulations. ZZ was partially supported by NSF grant AST-1208891. We gratefully acknowledge the use of the High Performance Computing Resource in the Core Facility for Advanced Research Computing at
Case Western Reserve University, the use of computing resources
at Shanghai Astronomical Observatory, and the support and resources from the
Center for High Performance Computing at the University of Utah.

The MultiDark Database used in this paper and the web application providing online access to it were constructed as part of the activities of the German Astrophysical Virtual Observatory as result of a collaboration between the Leibniz-Institute for Astrophysics Potsdam (AIP) and the Spanish MultiDark Consolider Project CSD2009-00064. The Bolshoi and MultiDark simulations were run on the NASA's Pleiades supercomputer at the NASA Ames Research Centre. The MultiDark-Planck (MDPL) and the BigMD simulation suite have been performed in the Supermuc supercomputer at LRZ using time granted by PRACE.

Funding for SDSS-III has been provided by the Alfred P. Sloan Foundation, the Participating Institutions, the National Science Foundation, and the U.S. Department of Energy Office of Science. The SDSS-III web site is http://www.sdss3.org/.

SDSS-III is managed by the Astrophysical Research Consortium for the Participating Institutions of the SDSS-III Collaboration including the University of Arizona, the Brazilian Participation Group, Brookhaven National Laboratory, University of Cambridge, Carnegie Mellon University, University of Florida, the French Participation Group, the German Participation Group, Harvard University, the Instituto de Astrofisica de Canarias, the Michigan State/Notre Dame/JINA Participation Group, Johns Hopkins University, Lawrence Berkeley National Laboratory, Max Planck Institute for Astrophysics, Max Planck Institute for Extraterrestrial Physics, New Mexico State University, New York University, Ohio State University, Pennsylvania State University, University of Portsmouth, Princeton University, the Spanish Participation Group, University of Tokyo, University of Utah, Vanderbilt University, University of Virginia, University of Washington, and Yale University.

\begin{appendix}
\section{Test of Fibre-Collision Correction Method with BOSS Ancillary
Data} \label{app:fiber}
\begin{figure*}
\includegraphics[width=0.8\textwidth]{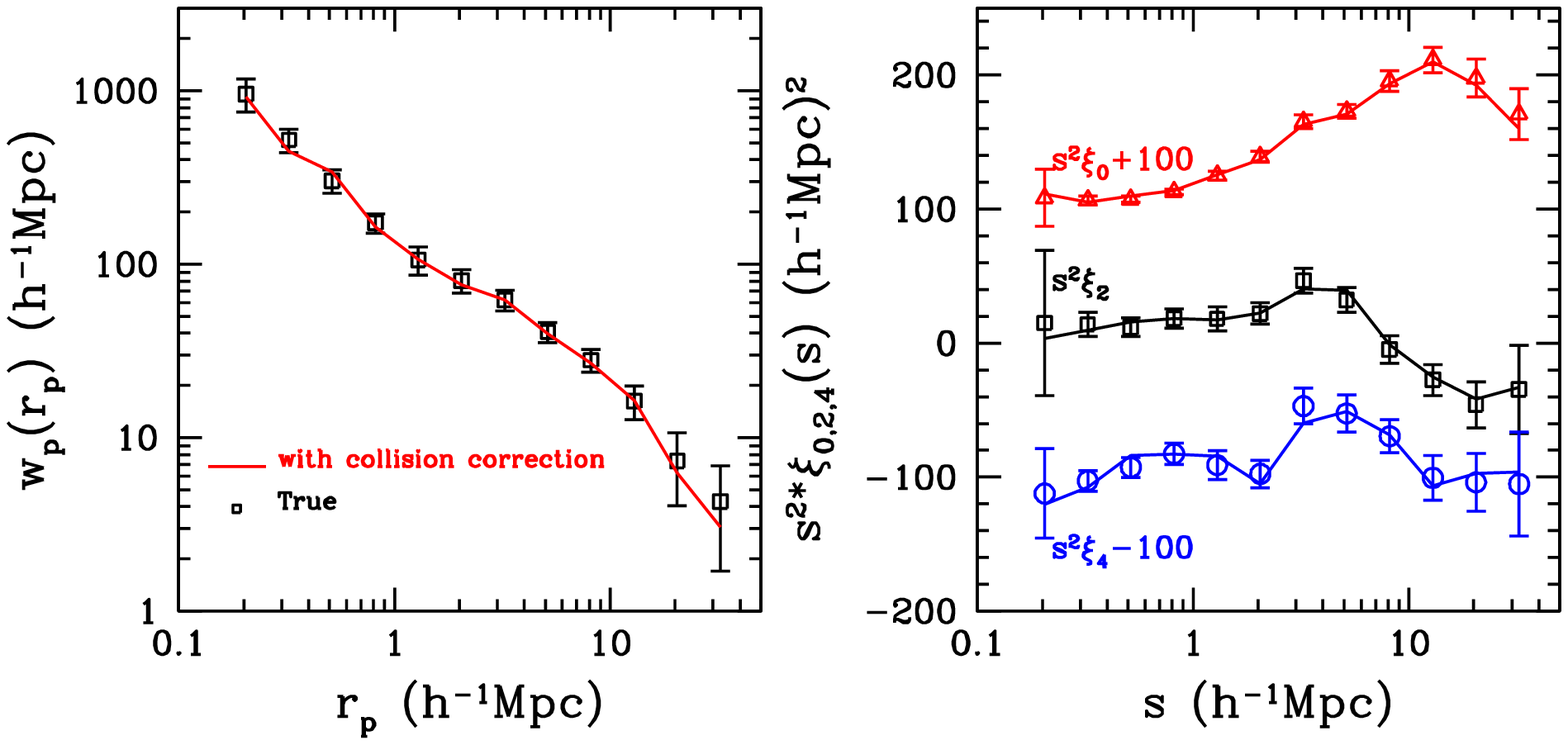}
\caption{Comparison between the true 2PCF measurements of the fibre-collision-free ancillary sample and those recovered from the fibre-collided sample by applying our fibre collision correction method. } \label{fig:fibercollision}
\end{figure*}
One of the BOSS ancillary programmes has targeted all the fibre-collided galaxies over a designated region of the survey, providing a fibre-collision-free sample of galaxies. Despite the small area of this ancillary sample and the smaller occurrence of fibre-collisions ($2.1$ per cent of the galaxies versus\  ${\sim}7$ per cent in the whole CMASS sample), it still provides an opportunity to test fibre-collision correction methods with observational data. Here we test the method of \cite{Guo12}, which is used in this paper. The method divides a source galaxy sample into two distinct populations, one free of fibre collisions and the other consisting of potentially collided galaxies. The total clustering signal is the appropriate combination of the contributions from these two populations, where the contribution of the collided population is estimated from the resolved galaxies in tile-overlap regions \cite[see details in][]{Guo12}.

Figure \ref{fig:fibercollision} shows the true measurements of the projected 2PCF $w_p(r_p)$ and the three multipole moments for the collision-free ancillary sample, compared to those recovered from the fibre-collided sample using the correction method of \cite{Guo12}. It is evident that within the error bars the fibre-collision method correctly recovers all the measurements in projected and redshift space in the ancillary sample. Farther tests relating to the robustness of our results to the fibre-collision correction were presented in Section 3.2.

\section{Modelling Results for the Full CMASS Sample}\label{app:full}
The full CMASS DR11 sample in the redshift range of $0.43<z<0.7$ covers a larger volume and is significantly larger than the volume-limited sample we utilize in this paper, and it has been used to constrain the cosmological parameters to $2$ per cent level by measuring the baryon acoustic oscillation feature \citep{Anderson13}. The complex sample definition of the full sample, however, make it non-trivial to be modelled by the simple HOD form. Nonetheless, we can still treat the 2PCF measurements as arising from an `average' sample at the median redshift and perform the HOD modelling. Given the importance of the full sample for cosmological constraints, in this appendix we present the SO halo modelling results of the full sample, including its constraints on velocity bias.

\begin{figure}
\includegraphics[width=0.4\textwidth]{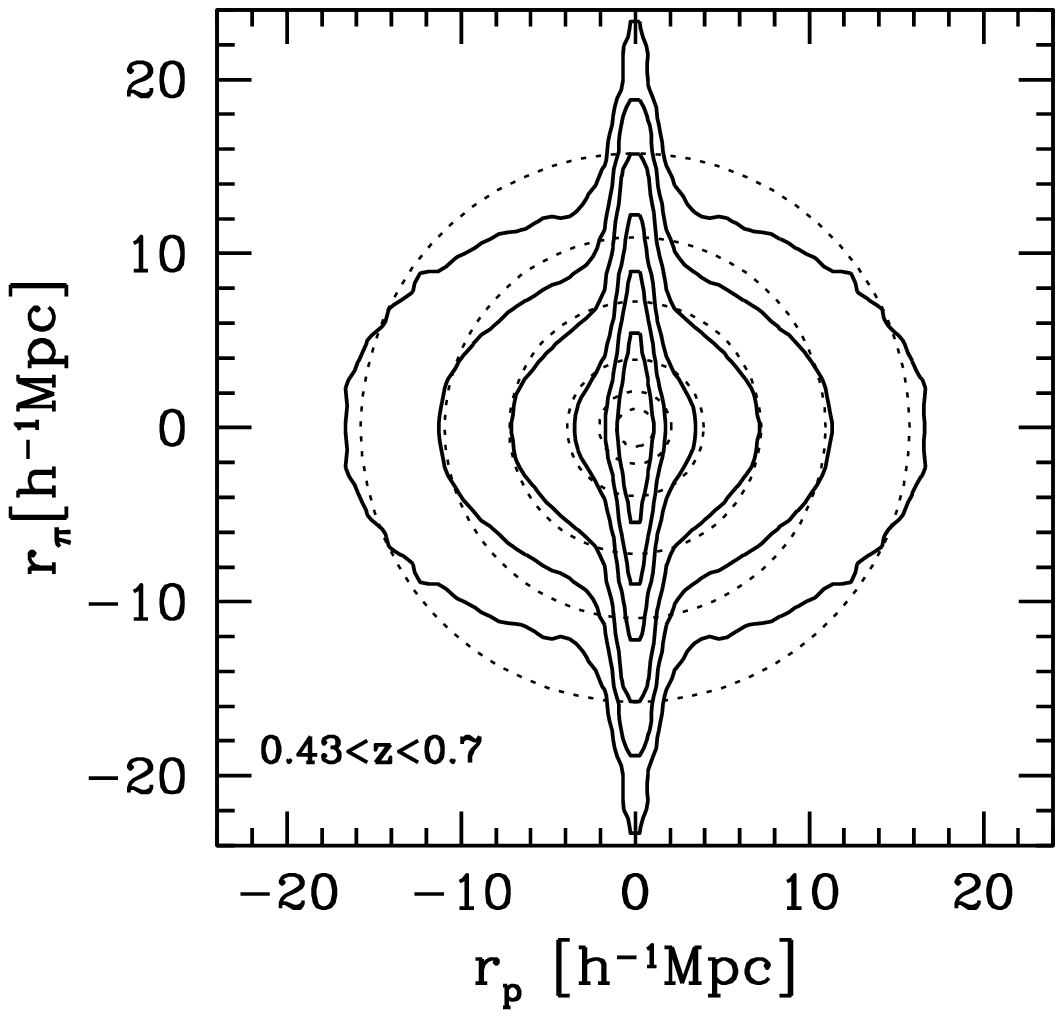}\caption{Measurements of the 3D 2PCF $\xi(r_p,r_{\rm\pi})$ for the for the full CMASS DR11 sample in the redshift range of $0.43<z<0.7$. Contour levels shown are $\xi(r_p,r_{\rm\pi})=[0.5, 1, 2, 5, 10, 20]$. The dotted curves are the angle-averaged redshift-space correlation function, $\xi_0(s)$, for the same sample and with the same contour levels. }\label{fig:xipfull}
\end{figure}
\begin{figure*}
\includegraphics[width=0.7\textwidth]{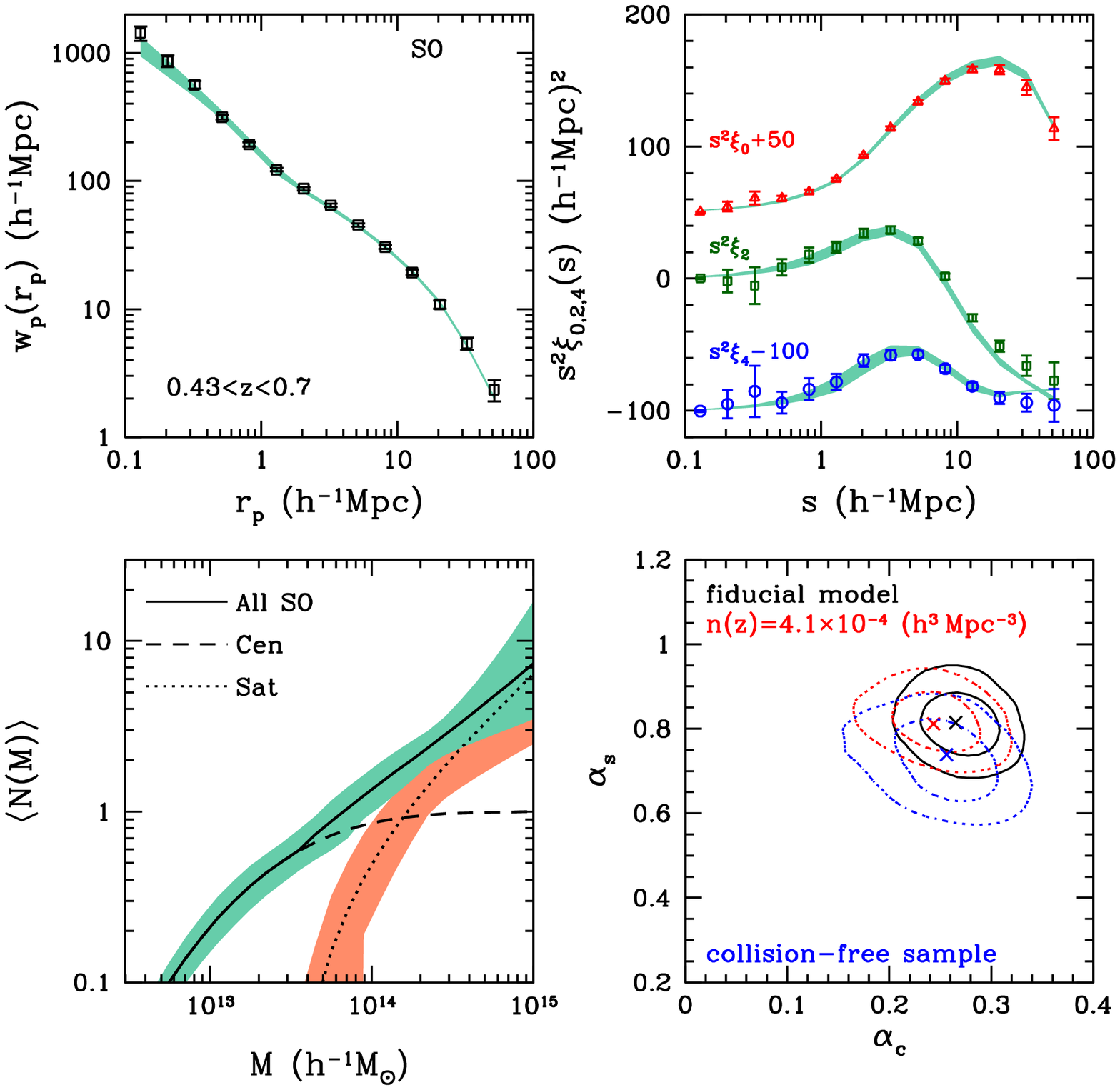}
\caption{ Measurements of the projected 2PCF and redshift-space 2PCF multipole moments and the HOD modelling results for the full CMASS sample. Top panels: similar to Fig.~\ref{fig:wpxi}, but for the full sample with the SO halo model. Bottom left panel: similar to Fig.~\ref{fig:hod}, but for the full sample. Bottom right panel: constraints on the galaxy velocity bias parameters for the full sample and robustness tests (see the text for details).} \label{fig:wpxifull}
\end{figure*}
\begin{table}
\caption{HOD parameters and satellite fraction derived for the full CMASS sample} \label{tab:full}
\begin{tabular}{lrr}
\hline
Parameters     & SO halo model & FOF halo model\\
\hline
$\chi^2/\rm{dof}$  & $68.14/50$ & $71.26/50$\\
$\log M_{\rm min}$ & $13.43^{+0.08}_{-0.06}$ & $13.42^{+0.07}_{-0.07}$\\
$\sigma_{\log M}$  & $0.75^{+0.04}_{-0.04}$  & $0.76^{+0.04}_{-0.04}$\\
$\log M_0$         & $13.54^{+0.15}_{-0.21}$ & $13.63^{+0.22}_{-0.24}$\\
$\log M_1^\prime$  & $14.08^{+0.11}_{-0.12}$ & $14.00^{+0.13}_{-0.27}$\\
$\alpha$           & $0.89^{+0.30}_{-0.25}$  & $1.20^{+0.28}_{-0.41}$\\
$\alpha_c$         & $0.27^{+0.03}_{-0.02}$  & $0.29^{+0.02}_{-0.02}$\\
$\alpha_s$         & $0.81^{+0.05}_{-0.05}$  & $0.64^{+0.04}_{-0.04}$\\
$f_{\rm{sat}}(\rm{per\,cent})$ & $5.99^{+0.35}_{-0.66}$ & $5.92^{+0.44}_{-0.63}$\\
\hline
\end{tabular}

\medskip
The mean number density of the full sample $\bar{n}(z)$ in $0.43<z<0.7$ is $2.17\times10^{-4}h^{3}{\rm {Mpc}}^{-3}$.
\end{table}
Fig.~\ref{fig:xipfull} shows the 3D 2PCF $\xi(r_p,r_{\rm\pi})$ for the full sample. Compared to the volume-limited $M_i<-21.6$ sample in Fig.~\ref{fig:xip}, the full sample has a slightly weaker FOG effect. This is a reflection of the fact that the full sample includes galaxies of lower luminosity and bluer colour, both of which have slower motions in haloes, leading to a weaker FOG feature. On large scales, the full sample has a stronger Kaiser effect, which can also be attributed to the inclusion of galaxies of lower luminosity and bluer colour, that have lower galaxy bias factors.

We present in Fig.~\ref{fig:wpxifull} and Table~\ref{tab:full} the modelling results for the full sample in the redshift range of $0.43<z<0.7$. The larger volume ($3.4\,h^{-3}{\rm Gpc}^3$) leads to tighter constraints on the velocity bias parameters, as shown by the solid black contours in the bottom right panel of Fig.~\ref{fig:wpxifull}. Compared to the volume-limited $M_i<-21.6$ sample, the velocity bias constraints tend to shift to slightly higher $\alpha_c$ and lower $\alpha_s$, but are still consistent with those from the $M_i<-21.6$ sample.

To investigate the robustness of the constraints of velocity bias in the full sample, we present the results of two additional tests in the bottom right panel of Fig.~\ref{fig:wpxifull}. One assumes a higher number density of the full sample, $\bar{n}(z)=4.1\times10^{-4}h^{3}{\rm {Mpc}}^{-3}$, as suggested by \cite{Reid14}, i.e. the observed full sample is regarded as a random subsample of a larger `parent' sample. This test is nearly equivalent to the case of keeping the original number density but setting a less-than-unity asymptotic value of the high-mass end central occupation function, as done with our volume-limited sample (\S~\ref{sec:sys}). The results are also similar, leading only to a slight shift in the central velocity bias (dotted red contours).

The other test is to constrain the velocity bias using the full sample limited to the `complete-sector sample' that is free of fibre collisions. The results here are also similar to those of the analogous test done with the volume-limited sample (\S~\ref{sec:sys}); we do not find substantial changes in the velocity bias constraints (dotted blue contours).

Recently, \cite{Reid14} performed modelling of the small- to intermediate-scale anisotropic clustering measurement for the full sample of CMASS DR10 galaxies to constrain $f\sigma_8$. They model modified monopole and quadrupole moments,
which effectively remove pairs in the FOG regime and thus slightly weaken the constraining power on the velocity bias compared to our modelling. In their one case with velocity bias constraints, they infer $\alpha_c=0.06\pm0.05$, consistent with no central velocity bias. Most of the discrepancy with our results can be accounted for by the use of different reference frames to define the velocity bias. While we use the inner 25 per cent particles around the potential minimum to define the core, they use the inner 3.75 per cent particles. There is a relative motion between the cores defined in these two ways \citep{Behroozi13,Reid14}. Accounting for this, we find that the two results for the central galaxy velocity bias are fully consistent. In other words, in a common frame, the centre-of-mass frame of
haloes, both become $\alpha_c\sim 0.31$.

The choice of the inner 3.75 per cent of particles in \citet{Reid14} (about 6 per cent of $R_{\rm vir}$) may serve as a guess for using dark matter structures (in dark matter only simulations) to represent central galaxies. However, realistically, the small central region is likely affected by baryonic processes, which makes the connection between dark matter only structures and galaxies less evident. For our choice of 25 per cent of the particles (about 27 per cent of $R_{\rm vir}$), the core is less sensitive to baryonic effects. Another advantage of this choice is that such a core can be well identified even in low resolution simulations, which is useful for modelling galaxy clustering in large redshift surveys (like BOSS and future extensions). While, for simplicity, one can define the velocity bias in the centre-of-mass frame of the haloes, defining a core component provides more information on the relation between central galaxies and the dark matter distribution.

\end{appendix}
\end{document}